\documentclass{emulateapj}
\usepackage{graphicx}
\usepackage[usenames]{color}
\usepackage{natbib}
\usepackage[normalem]{ulem}
\def \gtw{\>\hbox{\lower.25em\hbox{$\buildrel >\over\sim$}}\>}
\def \pccm3{pc-cm$^{-3}$}
\def \DDM{$\delta$(DM)}
\def \DM{{\rm DM}}
\slugcomment{Revised, 25 August 2016} 
\shorttitle{Single Pulses from the Crab Pulsar}
\shortauthors{Hankins, Eilek \& Jones}

\begin{document}

\title{The Crab Pulsar at Centimeter Wavelengths II: Single Pulses}
\author{T.\ H.\ Hankins\altaffilmark{1,2},  J.\ A.\ Eilek\altaffilmark{1,2}, G.\ Jones\altaffilmark{3,4}}
\altaffiltext{1}{Physics Department, New Mexico Tech, Socorro, NM 87801}
\altaffiltext{2}{Adjunct Astronomer, National Radio Astronomy Observatory}
\altaffiltext{3}{Jansky Fellow, National Radio Astronomy Observatory, Socorro, NM 87801}
\altaffiltext{4}{Columbia University, New York, NY 10027}
\email{thankins@aoc.nrao.edu}

\begin{abstract}
We have carried out new, high-frequency, high-time-resolution observations of the Crab pulsar.  Combining these with our previous data, we characterize bright single pulses associated with the Main Pulse, both the Low-Frequency and High-Frequency Interpulses, and the two High-Frequency Components.  Our data include observations at frequencies ranging from 1 to 43 GHz with time resolution down to a fraction of a nanosecond.  We find at least two types of emission physics are operating in this pulsar.  Both Main Pulses and Low-Frequency Interpulses,  up to $\sim$10 GHz, are characterized by \emph{nanoshot emission} -- overlapping clumps of narrow-band nanoshots, each with its own polarization signature. High-Frequency Interpulses, between 5 and 30 GHz, are characterized by \emph{spectral band emission} -- linearly polarized emission containing $\sim$30 proportionately spaced spectral bands.  We cannot say whether the longer-duration High-Frequency Component pulses are due to a scattering process, or if they come from yet another type of emission physics.
\end{abstract}


\section{Introduction} 
\label{sec:Intro}

The pulsar in the Crab Nebula does not play by the rules set down for ``normal'' pulsars.  Most pulsars have only one radio pulse per rotation period, and a few have two, but {\em seven} components have been found so far in the Crab's mean profile.  In many pulsars the radio pulse appears at a different rotation phase than the high-energy pulses, but in the Crab the two main radio pulses appear at the same phases as their high-energy counterparts. 

To make matters more interesting, the seven radio components do not behave as expected.  The star's mean radio profile is a strong function of frequency. Some components disappear, others appear, as one moves from low to high radio frequencies.  One might think one was looking at a totally different pulsar when observing below or above a few GHz. Furthermore, the temporal and spectral characteristics of individual radio pulses change dramatically among components.  This suggests that different physical conditions are operating in the regions which emit some of the components -- an idea that challenges symmetries built into all current models of the pulsar magnetosphere.

In previous work, our group studied the mean profile of the Crab pulsar at radio frequencies between $\sim 1$ and $8$ GHz \citep{MH1996, MH1999}, and extended that work to frequencies up to $\sim 43$ GHz \citep{Hankins2015}.  We carried out high-time resolution observations of individual bright pulses at frequencies $\sim 1$ to $10$ GHz \citep{Hankins2003,HE2007,Cross2010}.  In this paper we continue our single-pulse studies, extending our previous work to include polarization,  higher radio frequencies and more components of the Crab's mean profile.

\subsection{Components of the mean profile}

Each component in the mean profile of the Crab pulsar evolves with observing frequency \citep{MH1996,Hankins2015}. Two bright components dominate the mean profile at low radio frequencies (below a few GHz):  the  Main Pulse and the Low-Frequency Interpulse.  The latter lags the Main Pulse by $145^{\circ}$ of phase.  Two weaker components, also detectable below a few GHz, lead the Main Pulse by $36^{\circ}$ (the Low-Frequency Component) and by $20^{\circ}$ (the Precursor).  At high radio frequencies (between 9 and 30 GHz) the profile is dominated by three quite different components. Both the Main Pulse and the Low-Frequency Interpulse become weak and disappear altogether.  A new component, the High-Frequency Interpulse, leads the Low-Frequency Interpulse by $\sim 7^{\circ}$.  In addition, two new High-Frequency Components appear.  At 5 GHz they follow the High-Frequency Interpulse by $\sim 75^{\circ}$ and $\sim 130^{\circ}$ of phase, and their phase lag increases at higher observing frequencies.   We summarize these components in Table \ref{table:Components}. 

\begin{deluxetable}{llcc}
\tablecaption{Components of the Mean Profile \label{table:Components}}
\tablehead{
\colhead{Component} & \colhead{Acronym}& \colhead{Frequency} \\
                    &                  & \colhead{Range}     
}
\tablecolumns{3}
\startdata
Precursor                  & PC    & 0.3 -- \phn0.6 GHz\\
Main Pulse                 & MP    & 0.3 -- \phn4.9\phn GHz\\
High-Frequency Interpulse  & HFIP  & 4.2 -- 28.4 GHz\\
Low-Frequency Interpulse   & LFIP  & 0.3 -- \phn3.5 GHz\\
High-Frequency Component 1 & HFC1  & 1.4 -- 28.0 GHz\\
High-Frequency Component 2 & HFC2  & 1.4 -- 28.0 GHz\\
Low-Frequency Component    & LFC   & 0.6 -- \phn4.2 GHz\\[-2ex]
\enddata
\tablecomments{Frequency range over which component is detected in mean profiles.  Occasional single pulses may be detected outside this range, but they are too rare to contribute to the mean profile. From \citet{Hankins2015}.}
\end{deluxetable}

\subsection{Low or high emission altitudes?}
\label{Altitudes}

Pulsar radio emission has long been held to come from low altitudes close to the star's polar caps.  In the Crab pulsar, however,  the Main Pulse and the two Interpulses appear at the same rotation phases as the two bright components in the high-energy\footnote{In this paper we refer to optical, X-ray and $\gamma$-ray emission as ``high-energy'' emission.} light curve \citep{Abdo2010}. This is also the case for some millisecond pulsars \citep[e.g.,][]{Johns16}. Pulsed high-energy emission is now understood to come from extended ``caustic'' regions at  moderate to high altitudes in the magnetosphere \citep[e.g.,][]{Abdo2013}.  If the radio pulses were to come from the surface, they would lead the high-energy pulses by a substantial fraction of the rotation period \citep[e.g.,][]{Pierb16}.  For the Crab, the phase agreement of radio and high-energy pulses suggests the relevant emission regions are in the same parts of the magnetosphere.  In addition, our $\sim 60^{\circ}$ viewing angle relative to the star's rotation axis \citep{NgR04} precludes our seeing more than one bright polar-cap pulse per rotation period.  We therefore argue that \emph{radio emission from the Main Pulse and Interpulses of the Crab pulsar originates at high altitudes}. In \citet{EH2016} we suggest that the emission zones may lie between $(0.5 - 1.0)R_{\rm LC}$, where $R_{\rm LC}$ is the radius of the light cylinder.

The origin of the two High-Frequency Components is less clear.  Their unusual rotation phases suggest to us that they also come from high altitudes in the magnetosphere, but not from the same regions that produce the Main Pulse or the Interpulses.  The story of the Precursor and Low-Frequency Component may be different, however.  Because they appear earlier in rotation phase than the Main Pulse, and they do not have clear high-energy counterparts, it may be that they come from low altitudes, above one of the star's polar caps.  If this is the case, the star's magnetic axis must be inclined $\sim 60^{\circ}$ to its rotation axis, in order for our sight line to intersect a polar cap once per rotation period.
   
\subsection{Single pulses and emission physics}

In previous work we studied individual Main Pulses and High-Frequency Interpulses between 1 and 10 GHz \citep{Hankins2003,HE2007,Cross2010,Hankins2015}. We showed that Main Pulses have very different characteristics from High-Frequency Interpulses when observed at sufficiently high time resolution. Main Pulse emission is bursty on sub-microsecond timescales.  The spectrum of each ``microburst'' fills our observing band.  Once in awhile a Main Pulse is resolved into well-separated ``nanoshots,'' each lasting no more than a few nanoseconds.  High-Frequency Interpulse emission is also bursty, on few-microsecond timescales.  Its spectrum, however, is very different.  Emission from the High-Frequency Interpulse is confined to proportionately spaced \emph{spectral emission bands}, with band separation 6\% of the center frequency. 

We believe the very different characteristics of Main Pulses and High-Frequency Interpulses show that different physics is operating in the parts of the magnetosphere which lead to each component. We recall that three conditions must converge to create pulsar radio emission \citep[see also][]{EH2016}.  (1) There must be a source of available energy which can be tapped for radio emission. This may be relativistic plasma flow in the open field line region;  magnetic reconnection has also been suggested.  (2) There must be a mechanism by which the available energy is converted to coherent radio emission.  This probably involves coherent charge motion and may start with a plasma instability. (3) Finally, there must be a long-lived site in the magnetosphere within which conditions are right for items (1) and (2) to work.  The polar caps provide one such region;  high-altitude caustics may be another. 

We collectively refer to all of these requirements as the ``emission physics'' responsible for pulsed, coherent radio emission.  We believe our high-time-resolution observations provide important clues to the emission physics for mean-profile components of the Crab pulsar.  In this paper we present new data for Main Pulses, both Interpulses and the High-Frequency Components, and we briefly discuss possible emission physics for each of these components.  In a companion paper, \citet{EH2016}, we critique models of the radio emission mechanism and use our data to constrain conditions in the emission zones for the bright components. 

\subsection{What's coming in this paper}

In this paper we present new single-pulse data between 14 and 43 GHz, and combine it with our previous data below 10 GHz to characterize the five bright components of the Crab pulsar's mean profile. 

We begin with the Main Pulse and the Low-Frequency Interpulse.  After an overview of our observations, in Section \ref{section:observations}, we discuss the short-lived microbursts which characterize Main Pulses in Section \ref{section:MainPulses}. We show that microbursts continue in Main Pulses up to 22 GHz, and that Low-Frequency Interpulses have the same temporal and spectral signature as Main Pulses. In Section \ref{section:InternalToMicrobursts} we explore the polarization of Main Pulses down to few-nanosecond time resolution.  We show that polarization in Main Pulses can fluctuate on such short timescales, and argue that both Main Pulses and Low-Frequency Interpulses are composites of overlapping nanoshots.  In Section \ref{section:QbandMainPulse} we present one very interesting Main Pulse we captured at 43 GHz, which suggests interesting new physics at this high frequency.

We then move to the High-Frequency Interpulse. In Section \ref{section:IP_band_structure} we present new, high-frequency observations of the spectral emission bands. We show the proportional frequency spacing of the bands continues up to 30 GHz, and we critique some models which have been proposed to explain the bands.  In Section \ref{section:IP_Polarization} we show that individual High-Frequency Interpulses show strong linear polarization with constant position angle that is independent of the rotation phase at which the pulse appears. In Section \ref{section:IP_Dispersion} we show that the High-Frequency Interpulse is characterized by variable, intrinsic dispersion that must arise locally to the emission region for that component.

In Section \ref{section:HFCs} we turn to the two High-Frequency Components, and present single pulses in each component that we captured at 9 GHz. We show that individual High-Frequency-Component pulses are relatively weak, but last much longer, than typical Main Pulses or Interpulses.  We consider possible origins of these two components, and ask whether they represent yet another type of emission physics in this pulsar. 

Finally, in Section \ref{section:Last}, we summarize our results, discuss what we have learned about this pulsar, and point out questions which still need answers. 

\section{Observations} 
\label{section:observations}

We recorded the data we present in this paper during several observing sessions at the Karl G.\ Jansky Very Large Array,\footnote{The Very Large Array (VLA) is an instrument of the National Radio Astronomy Observatory, a facility of the National Science Foundation operated under cooperative agreement by Associated Universities, Inc.} 
at the Arecibo Observatory,\footnote{The Arecibo Observatory (AO) is operated by SRI International under a cooperative agreement with the National Science Foundation (AST-1100968), and in alliance with Ana G.\ M\'endez-Universidad Metropolitana, and the Universities Space Research Association.}
at the Robert C.~Byrd Green Bank Telescope,\footnote{The Green Bank Telescope (GBT) is an instrument of the National Radio Astronomy Observatory, a facility of the National Science Foundation operated under cooperative agreement by Associated Universities, Inc.} 
and at the Goldstone-Apple Valley Radio Telescope.\footnote{The Goldstone-Apple Valley Radio Telescope (GAVRT) is operated by the Lewis Center for Educational Research with support from the NASA Jet Propulsion Laboratory.} 

We used our Ultra High Time Resolution System (UHTRS), as described in \protect{\citet{Hankins2015}} to capture individual pulses from the Crab pulsar.  In this system, a digital oscilloscope, triggered by pulses that exceed six times the noise levels as tabulated in Table \ref{table:Noise_Levels}, sampled and recorded the received voltages from both polarizations.  
The data were processed off-line using coherent dedispersion \citep{Hankins1971,HR1975}, which allowed time resolution down to the inverse of the receiver bandwidth.

 Because pulses bright enough for us to record with the UHTRS formally coincide with the high-flux tail of the number versus flux histogram for single pulses, as seen by \citet{Argyle1972} and \citet{Lundgren1995} at lower frequencies, they might loosely be called ``giant'' pulses. However, it is not yet clear whether such high-flux pulses are physically similar to, or different from, the more common ``weak'' pulses.  In fact, the distribution of pulse amplitudes in the Crab pulsar appears to be continuous from weak to strong \cite[see, e.g.,][]{Karuppusamy2010}.  In what follows -- as also in \citet{HE2007,Cross2010,Hankins2015} -- we do not attempt to distinguish between weak and strong pulse populations,  but just discuss Main Pulses, Interpulses and High-Frequency-Component Pulses.
 
 We have included relevant telescope and observing parameters in Appendix I.

\begin{figure}[htb]  
\center{
\includegraphics[width=\columnwidth]{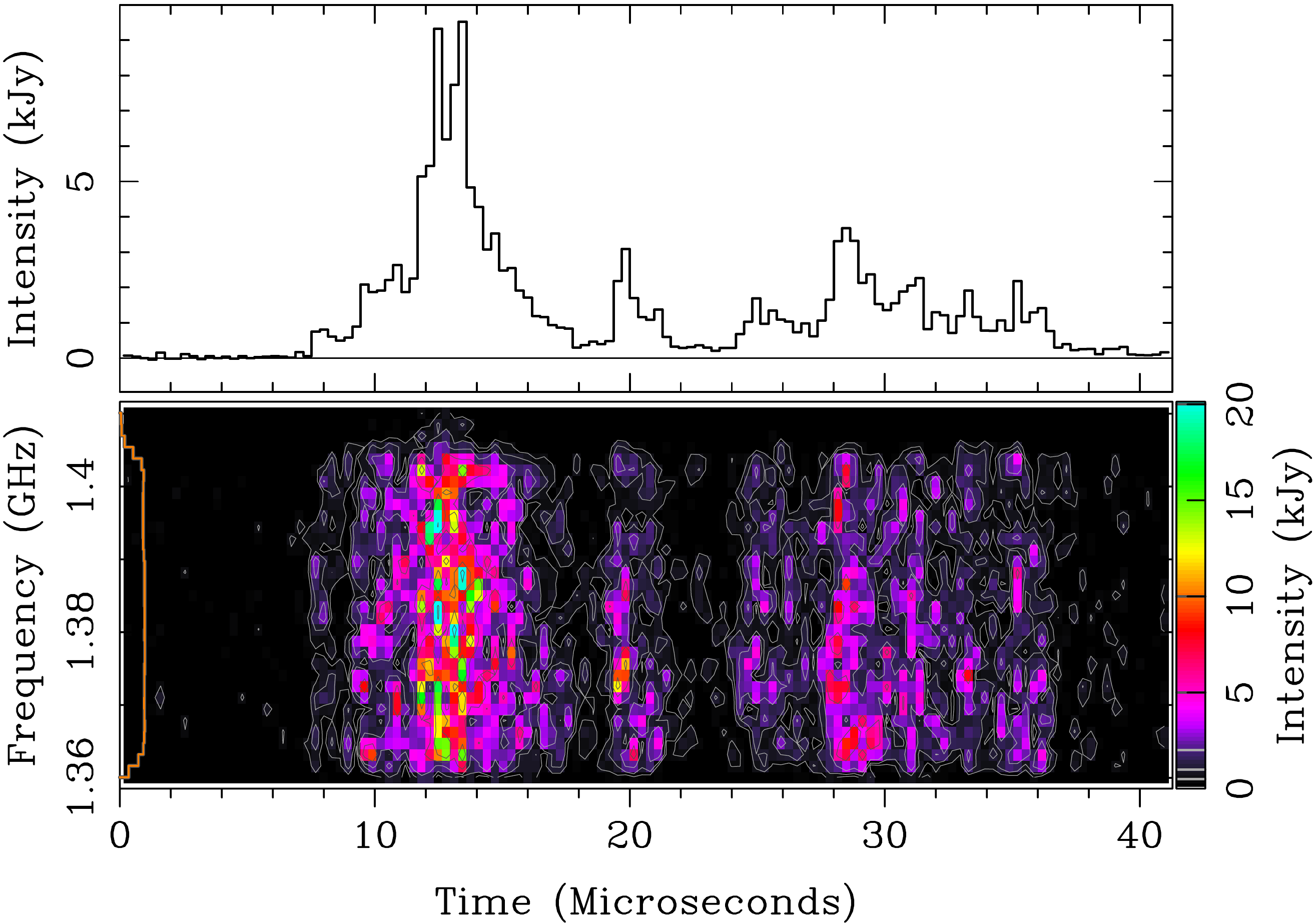}
\includegraphics[width=\columnwidth]{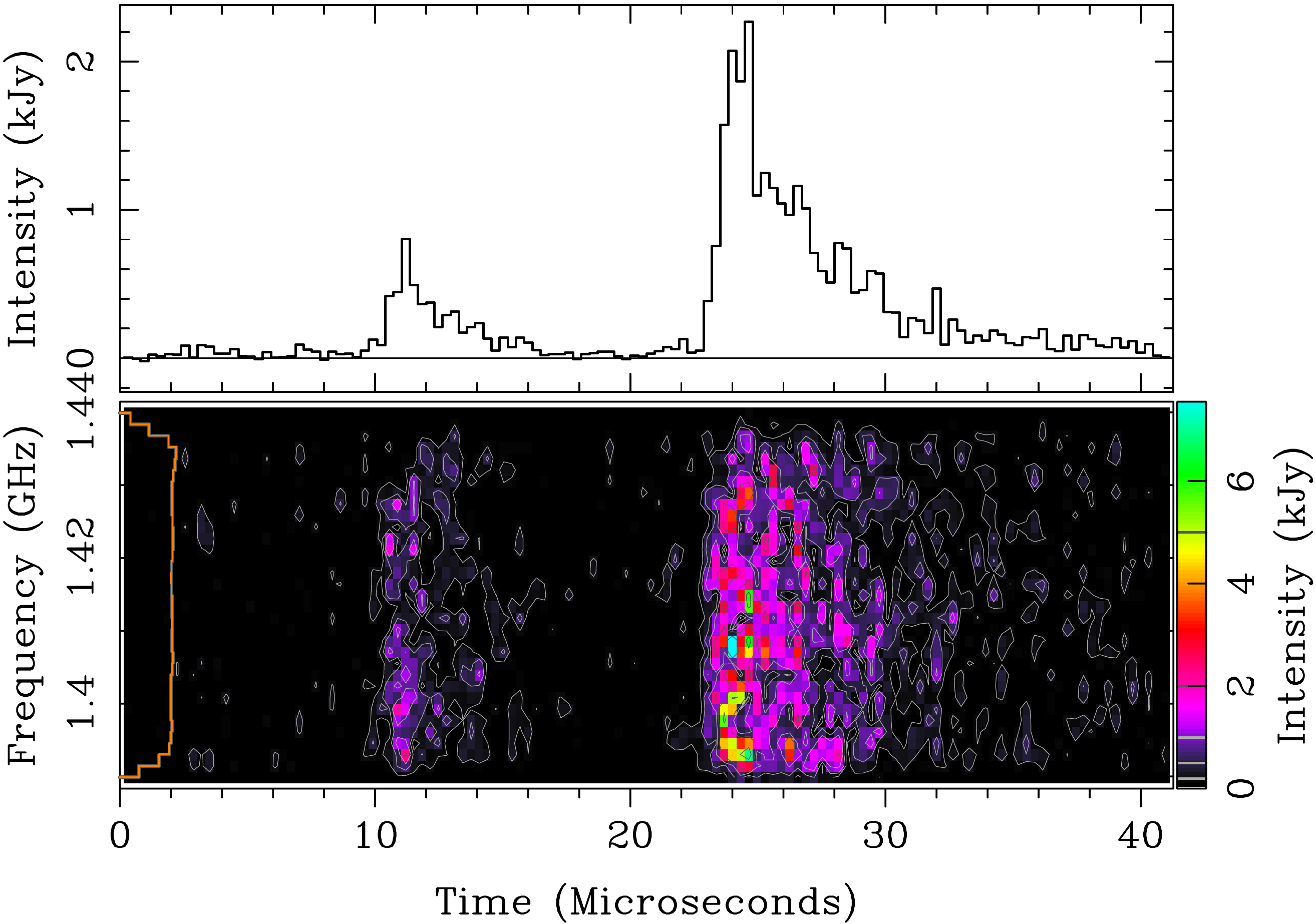}
\caption[]{Comparison of a  Low-Frequency Interpulse and a Main Pulse: both contain microbursts with spectra extending across the observing band.  Top, a typical Low-Frequency Interpulse seen at 1.4 GHz,  plotted with 320 ns time resolution and 1.5625 MHz spectral resolution,  de-dispersed with dispersion measure (DM) of 56.8270 \pccm3. Contour levels are 0.5, 1, 2, 5, 10, and 20 kJy. Bottom, a typical Main Pulse at 1385 MHz, plotted at 320 ns time resolution, and 1.5625 MHz spectral resolution, de-dispersed with DM of 56.79682 \pccm3. Contour levels are 0.2, 0.5, 1, 2, 5 kJy. In this and later figures, the orange line to the left is the equalized off-pulse response function of the receiver.}
\label{fig:LFIPandLFMP}
}
\end{figure}

\section{Microbursts: variable energy supply?}
\label{section:MainPulses}

We previously found that Main Pulses, captured between 1 and 10 GHz, contain microsecond-long radio bursts \citep{HE2007,Cross2010,Jessner2010}.  Microbursts typically last a few microseconds,  and are shorter at higher frequencies.  They can occur anywhere in the probability envelope defined by the mean-profile component, which extends for several hundred microseconds \citep[e.g.,][]{Hankins2015}.  Their spectrum is relatively broadband:  it extends at least across the 2.5 GHz observing band we used in \citet{HE2007}.  In this section we extend our Main Pulse study to larger bandwidths, higher radio frequencies and polarization studies.  We also show that the Low-Frequency Interpulse has the same temporal and spectral characteristics as does the Main Pulse.  

\subsection{Microbursts in Low-Frequency Interpulses}

 Low-Frequency Interpulses and Main Pulses share similar temporal and spectral properties. In Figure \ref{fig:LFIPandLFMP} we show one example of a Low-Frequency Interpulse and a comparable Main Pulse, both observed near 1.4 GHz at the Very Large Array.  Although the Low-Frequency Interpulse is not detectable in the mean profile much above 4 GHz \citep{Hankins2015}, we have captured a few single Low-Frequency Interpulses at higher frequencies.  \citet{Jessner2010} also present one 8 GHz pulse which they captured at the rotation phase of the Low-Frequency Interpulse.  These data show that the temporal and spectral characteristics of  Low-Frequency Interpulses are the same as for Main Pulses.  The emission usually arrives in one or a few microsecond-long microbursts, the spectrum of which is continuous across the observing band.  We show in Section \ref{Nanoshots} that an occasional Low-Frequency Interpulse is found to  contain several well-separated nanoshots. Based on these similarities to characteristics of the Main Pulse,  we conclude that the same emission physics governs both components.

\subsection{Microbursts in Main Pulses}
\begin{figure}[htb] 
\includegraphics[width=\columnwidth]{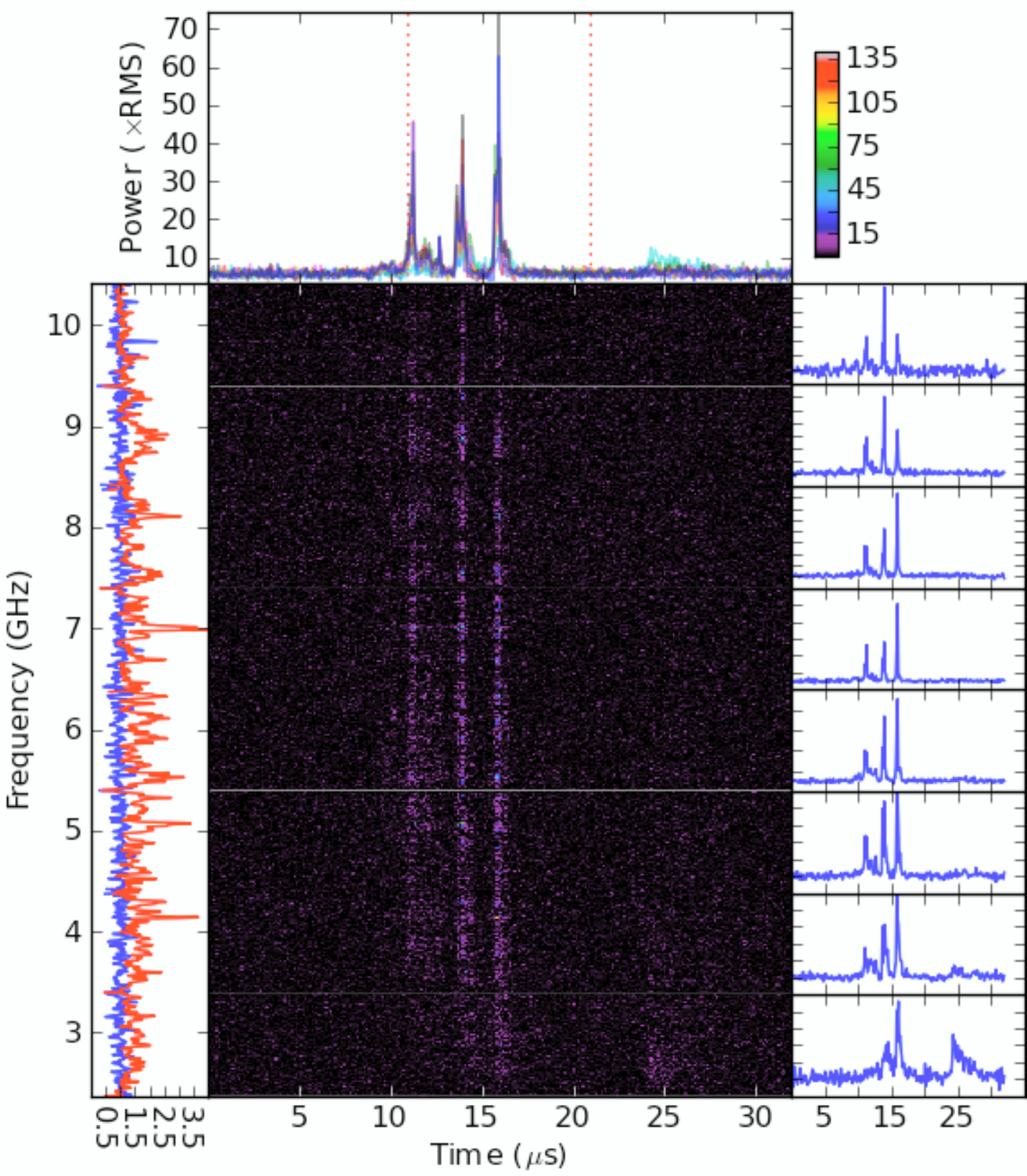}
\caption[]{A typical Main Pulse, with three microbursts, captured between 2 and 10 GHz with 10 ns time resolution using the  Goldstone-Apple Valley Radio Telescope. The spectrum resolution is 8 MHz. The color bar at the top right describes the power in the frequency-time plane. The left panel shows on-pulse (red) and off-pulse (blue) power as a function of frequency, integrated across the pulse.  The high spikes in the on-pulse power correspond to strong, narrow-band spikes within each microburst.  The right panel shows the pulse shape at 8 frequencies within the pulse.  All 8 pulses are over-plotted in the top figure, where the dotted lines separate the on-pulse and off-pulse regions used for the left panel. All powers given in terms of off-pulse noise (RMS).}
\label{fig:GAVRT_pulse}
\end{figure}

 To probe larger bandwidths, we used the GAVRT 34-m telescope to record single Main Pulses with an observing band extending from 2 to 10 GHz, and $10$ ns time resolution. We show one example of a multi-burst Main Pulse, captured by multiple CASPER\footnote{Collaboration for Astronomy Signal Processing and Electronics Research,  https://casper.berkeley.edu} iBob devices at GAVRT, in Figure \ref{fig:GAVRT_pulse}.  These results show that the microburst spectrum extends at least over the full 8 GHz frequency range in the pulses we captured with GAVRT. 

\begin{figure}[htb] 
\includegraphics[width=\columnwidth]{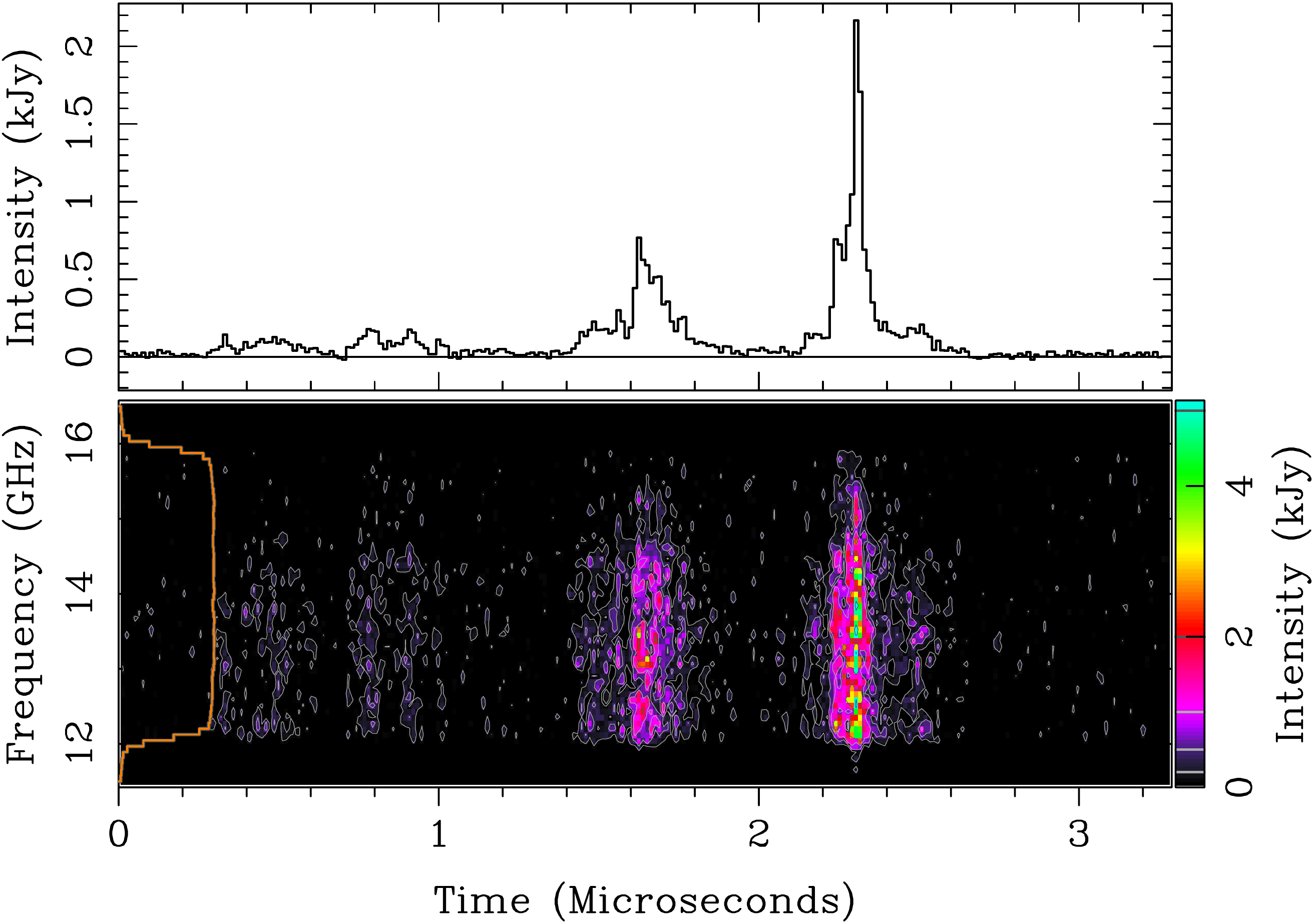}
\includegraphics[width=\columnwidth]{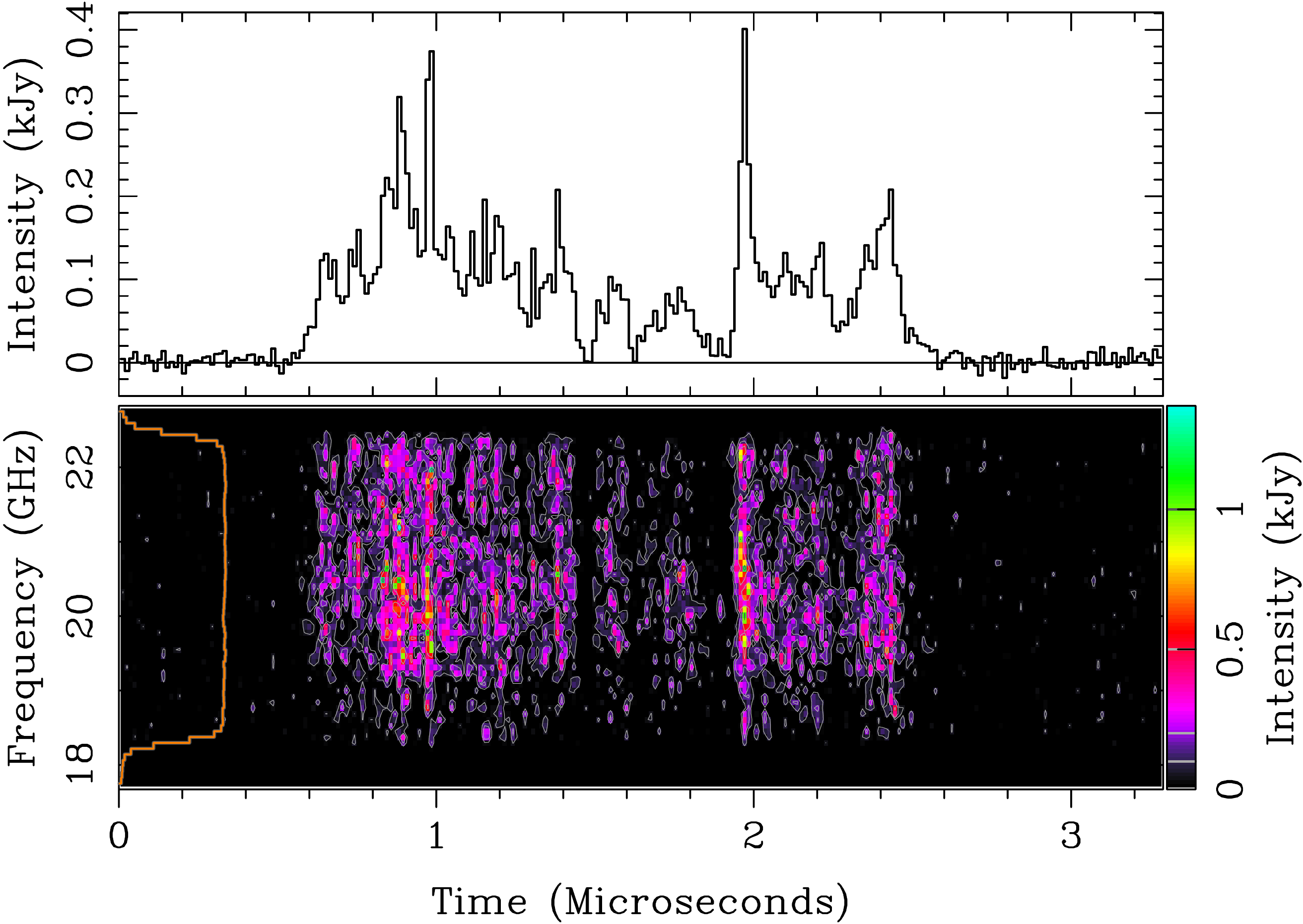}
\caption[]{Two Main Pulses recorded above 10 GHz.  In agreement with Main Pulses seen at lower frequencies \citep{HE2007}, these two examples contain a few microbursts, and have a spectrum extending across our observing band.  Top, a Main Pulse at $12-16$ GHz,  shown with 12.8 ns time resolution, 78 MHz spectral resolution, and de-dispersed with DM of 56.78900 \pccm3. The contour levels are 0.1, 0.2, 0.5 and 1 kJy. Bottom: a  Main Pulse at $18.3-22.5$ GHz, shown with time resolution 12.8 ns, spectral resolution 78 MHz, and de-dispersed with DM of 56.77800 \pccm3. The contour levels are 0.2, 0.5, 1 and 5 kJy. The microbursts here are narrower than those in Figure \ref{fig:LFIPandLFMP}, consistent with the width-frequency trend found in \citet{Hankins2015}. 
\label{fig:MPs_above_10GHz} 
}
\end{figure}

To probe higher frequencies, we recorded single pulses with our UHTRS between 14 and 43 GHz at the GBT.  As shown in Figures 4 and 5 of \citet{Hankins2015} at frequencies above about 10 GHz the occurrence of strong Main Pulses is rare as compared with strong High-Frequency Interpulses. In fact, using the UHTRS we captured only 6 Main Pulses above 10 GHz in several tens of observing hours. We show in Figure \ref{fig:MPs_above_10GHz} two Main Pulses, one at 14 and one at 20 GHz. It is clear that the nature of Main Pulses is the same above and below 10 GHz. The same microburst structure is still  apparent at these higher frequencies,  and the spectrum of each microburst continues across our full observing band.  We conclude that Main Pulses are well described as broadband microbursts across the entire 1 to 30 GHz frequency range where we have studied them. 

\subsection{A note on dispersion measure}
\label{DM_note}

At our high time resolution, details of the pulses are sensitive to the exact value of the dispersion measure (DM) used for the coherent dedispersion operation. To determine an ``optimum DM'' for the figures presented in this paper we generally started with the DM value given by the Jodrell Bank Crab Pulsar Monthly Ephemeris\footnote{See http://www.jb.man.ac.uk/$\sim$pulsar/crab.html} for each observing epoch.  From there, we followed one of two paths.  For sharp pulses, such as bursts within a Main Pulse, we calculated the intensity variance after dedispersion with a range of DM values around the Jodrell value and used the DM which maximized the variance. Alternatively, for broad pulses such as the High-Frequency Components and the High-Frequency Interpulse, we experimented with small DM changes to determine ``by eye'' the DM value that best aligned the start of the pulse in the dynamic spectrum. 
With both methods, we typically found it necessary to refine our ``optimum'' DM values up to $\sim 10^{-2}$ \pccm3 relative to the Jodrell value.

In Section \ref{section:IP_Dispersion} we report on systematic methods to determine the DM of single pulses that are more appropriate for analyzing a large number of pulses.  Our results in that section verify the size of the difference between the DM of individual pulses and the Jodrell value.   

\subsection{Sporadic energy release in the emission zones}

Microbursts show that the Main Pulse and Low-Frequency Interpulse emission zones are neither uniform nor homogeneous.  They must be variable in space and time.  We envision isolated, short-lived regions where energy builds up until it some or all of it is released in a burst of coherent radio emission. After a period of quiescence, perhaps the same region is re-energized, or perhaps the process moves to a different part of the emission zone.  One possibility, which we discuss in \citet{EH2016}, is that this sporadic energization involves a ``sparking'' cycle \citep{RS75} in an unsteady pair cascade.  The details are not yet clear, however, because it is not known how such a pair cascade cycle operates (if at all) at the high altitudes where the Main Pulse and Low-Frequency Interpulse originate.   

\section{Main Pulses: Polarization and Nanoshots}
\label{section:InternalToMicrobursts}

The Main Pulse and Low-Frequency Interpulse are only weakly polarized in the mean radio profile \citep{MH1999, Slo15}.   \citet{Slo09} also found low fractional polarization for the \emph{optical} Main Pulse and Interpulse, and suggested the cause is depolarization from extended emission in two-pole caustics \citep{Dyks04}.  Their suggestion is supported by rapid position angle swings in the two optical pulses,  and is consistent with a high-altitude caustic origin for optical emission.

 However, observations of single radio pulses tell a different story.  Weak polarization in those pulses is more complicated than simple caustic effects. Radio observations of single pulses by \citet{Hankins2003} with 2 ns time resolution,  \citet{Sog07} with $\sim 30$ ns time resolution, and \citet{Jessner2010} with 2 ns time resolution, show rapid changes of linear and circular polarization on time scales of a few nanoseconds. Such rapid variability depolarizes individual pulses observed at low time resolution, as well as the associated mean-profile components.

\subsection{Polarization suggests nanoshots}
Our larger sample of Main Pulses confirms and extends these results. In particular, we find that some -- but not all -- Main Pulses show rapid polarization fluctuations on a timescale of a few nanoseconds.  To simplify the discussion in this section, we use ``low time resolution'' to mean smoothing to $\sim 50-100$ ns, and ``high time resolution'' to mean smoothing to only a few nanoseconds.  Our best possible time resolution, of course, is the inverse of our observing bandwidth, i.e., a fraction of a nanosecond for most of the observations presented here.  To reduce uncertainty in the polarized flux, we always smooth our data by at least several times that limit. 

\begin{figure}[htb] 
\includegraphics[width=\columnwidth]{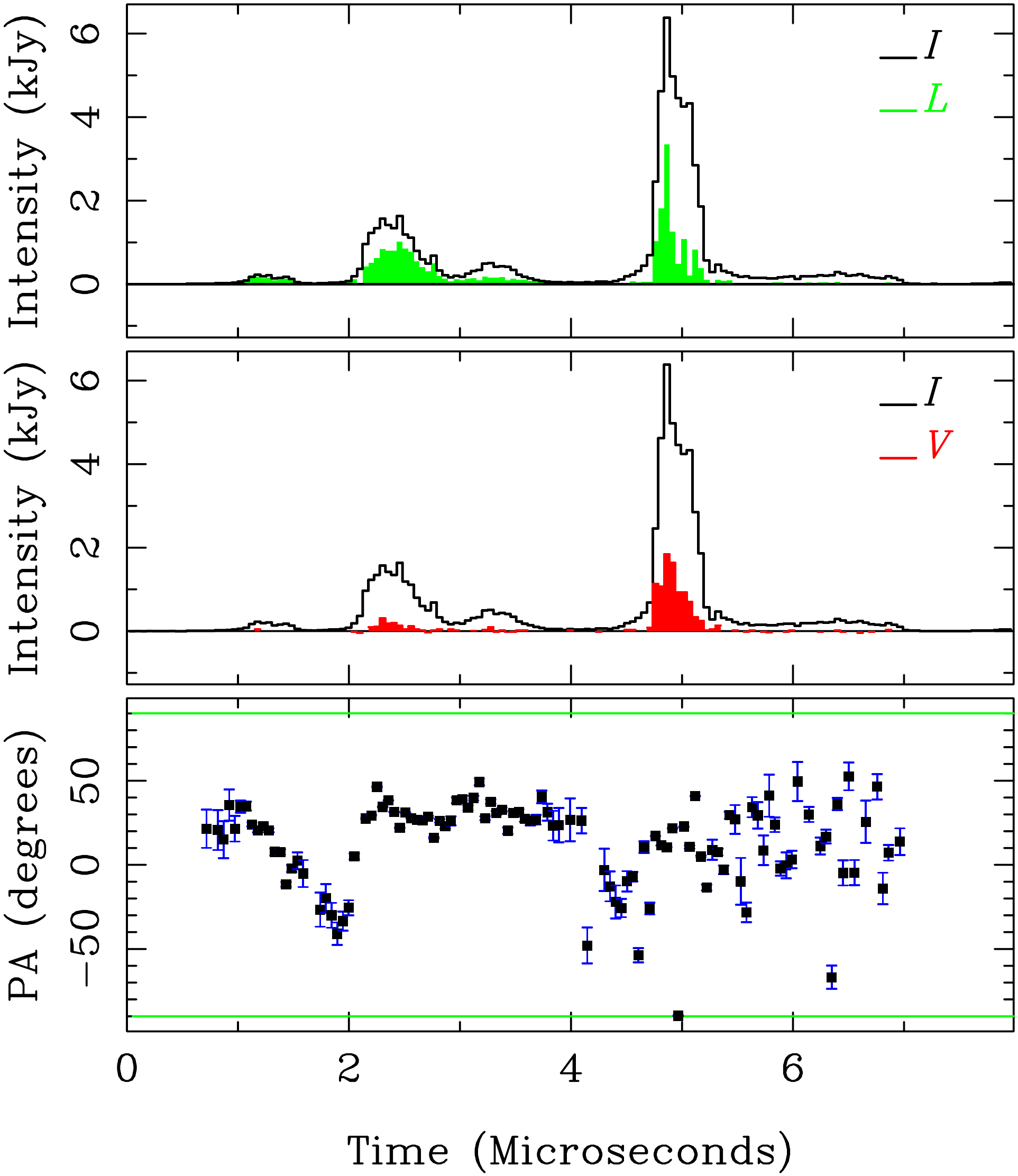}
\caption[]{Polarization of a Main Pulse,  captured between 8 and 10.5 GHz, de-dispersed with DM of 56.75252 \pccm3 and displayed with  51.2 ns time resolution. Top panel shows total ($I$; black) and linearly polarized ($L$; green) intensity. Middle panel shows total and circularly polarized ($V$;  red) intensity. Bottom panel shows position angle of linear polarization.  Orange lines -- indistinguishable from zero-intensity axis in this example but apparent in later figures -- indicate 4 times the off-pulse noise level;  only polarization above this threshold is shown.  In this pulse, the position angles are ordered in the first three microbursts, but become disordered in the strongest microburst and the lagging tail. Compare the polarization behavior to that of the pulses in Figures \ref{fig:MP_disorder_1}  and \ref{fig:MP_disorder_2}. }
\label{fig:MP_ordered_pol} 
\end{figure}

\subsubsection{Low time resolution: ordered polarization is rare}

Some Main Pules observed at low time resolution show strong linear polarization and ordered position angle behavior.  \citet{Jessner2010} showed two examples. In Figure \ref{fig:MP_ordered_pol} we show another, at 51.2 ns time resolution. The leading part of this pulse shows strong polarization and ordered position angle rotation.  However, the position angle behavior appears to change with each microburst -- clearly inconsistent with the simple rotating-vector model of \citet{RC69}. Interestingly, the position angle behavior of this pulse changes again in the last two microseconds, when it becomes apparently random. 

\begin{figure}[htb] 
\includegraphics[width=\columnwidth]{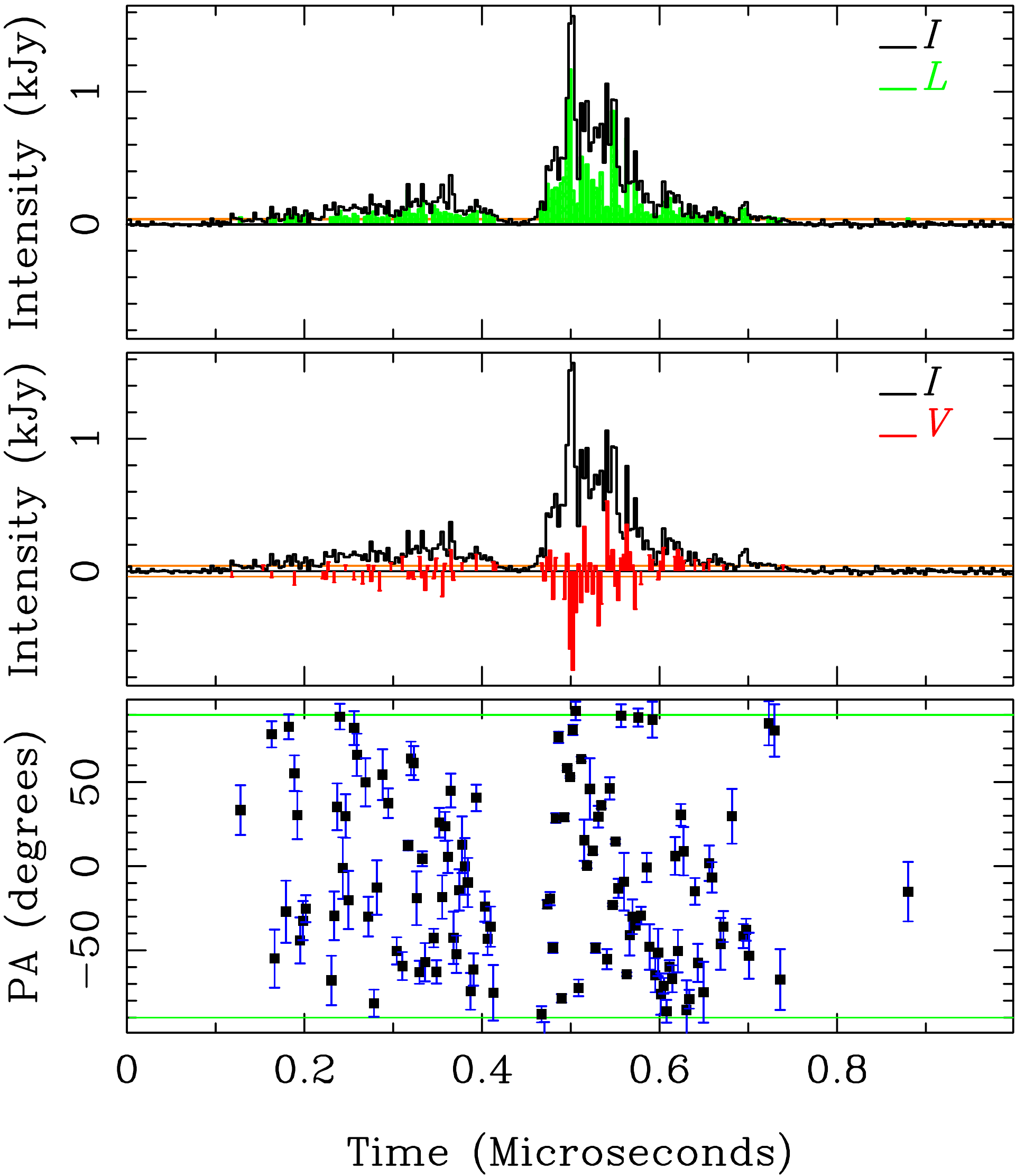}
\caption[]{Polarization of two microbursts in another Main Pulse, captured between 8 and 10.5 GHz, de-dispersed with DM of 56.76080 \pccm3, and displayed with 3.2 ns time resolution. Layout is the same as in Figure \ref{fig:MP_ordered_pol}.  Both the amount and the position angle of the linear polarization change rapidly in this pulse, as does the magnitude and sign of the circular polarization. Compare the polarization behavior to that of the pulses in Figures \ref{fig:MP_ordered_pol} and \ref{fig:MP_disorder_2}.}
\label{fig:MP_disorder_1} 
\end{figure}

Such disordered behavior is typical of most of the Main Pulses we have captured, where well-ordered position angle behavior is the exception rather than the rule.  At low time resolution, most Main Pulses have disordered position angle behavior, and linear polarization a small fraction of the total intensity. 

The story is similar for circular polarization.  Some Main Pulses show significant circular polarization at low time resolution, such as the pulse in Figure \ref{fig:MP_ordered_pol}.  In other examples circular polarization is significant, but changes sign between microbursts.  Still other  pulses show little or no circular polarization when smoothed to low time resolution.

\subsubsection{High time resolution:  polarization fluctuates rapidly}

The story becomes more interesting when pulses are studied at high time resolution.  Main Pulses we have captured often show significant linear and circular polarization which fluctuates  on timescales no longer than a few nanoseconds.

\begin{figure}[htb] 
\includegraphics[width=\columnwidth]{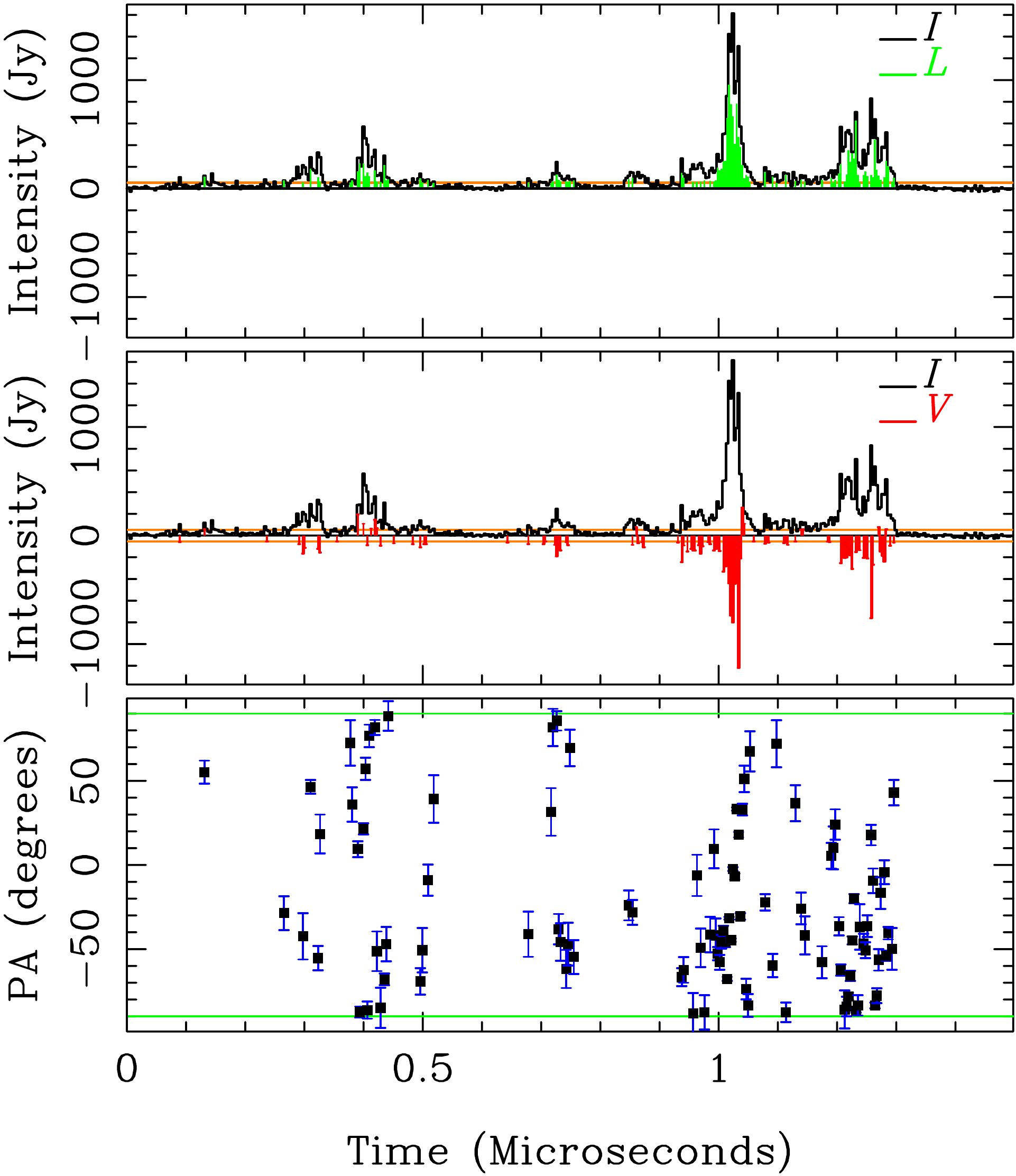}
\caption[]{Polarization of several microbursts in a third Main Pulse, captured between 8 and 10.5 GHz, de-dispersed with DM of 56.75900 \pccm3, and displayed with 3.2 ns time resolution. Layout is the same as in Figure \ref{fig:MP_ordered_pol}.   Both the intensity and the position angle of the linear polarization change rapidly in this pulse, similar to the pulse in Figure \ref{fig:MP_disorder_1}.  Circular polarization in this pulse fluctuates rapidly, but maintains the same sign throughout most of the pulse;  this behavior differs from the pulse in Figure \ref{fig:MP_disorder_1}, but resembles the pulse in Figure  \ref{fig:MP_ordered_pol}.  }
\label{fig:MP_disorder_2} 
\end{figure}

In Figures \ref{fig:MP_disorder_1}  and \ref{fig:MP_disorder_2} we show two typical examples displayed at 3.2 ns time resolution.   In both examples, the position angle of the linear polarization fluctuates on timescales of a few nanoseconds.  These  pulses would be only weakly polarized if studied at lower time resolution. Circular polarization in the pulse in Figure \ref{fig:MP_disorder_1} changes sign on similar timescales.  This pulse loses circular polarization when studied at lower time resolution.   By contrast, circular polarization of the pulse in Figure \ref{fig:MP_disorder_2} is predominantly negative, but rapid sign changes can be seen in the two leading microbursts.  

Figures \ref{fig:MP_disorder_1}  and \ref{fig:MP_disorder_2} also show strong, rapid fluctuations in total intensity.   Because these fluctuations are much larger than off-pulse noise, we believe they are physical.  We have argued before that  microbursts in the Main Pulse and the Low-Frequency Interpulse are not the smallest unit of emission.  Our previous high time resolution observations show that Main Pulses can occasionally be resolved into shorter-lived structures, which we have called ``nanoshots''  \citep[][see also Jessner et al. 2010]{Hankins2003,HE2007}. Such rapid total intensity fluctuations are very suggestive of blended nanoshots, in agreement with the amplitude-modulated noise model of \citet{R75}.  Our polarization data suggest that each nanoshot has its own polarization identity -- its own linear and circular polarization signature -- which may or may not change from one nanoshot to the next.  

\subsection{Nanoshots observed directly}
\label{Nanoshots}

\begin{figure}[htb] 
\center{
\includegraphics[width=\columnwidth]{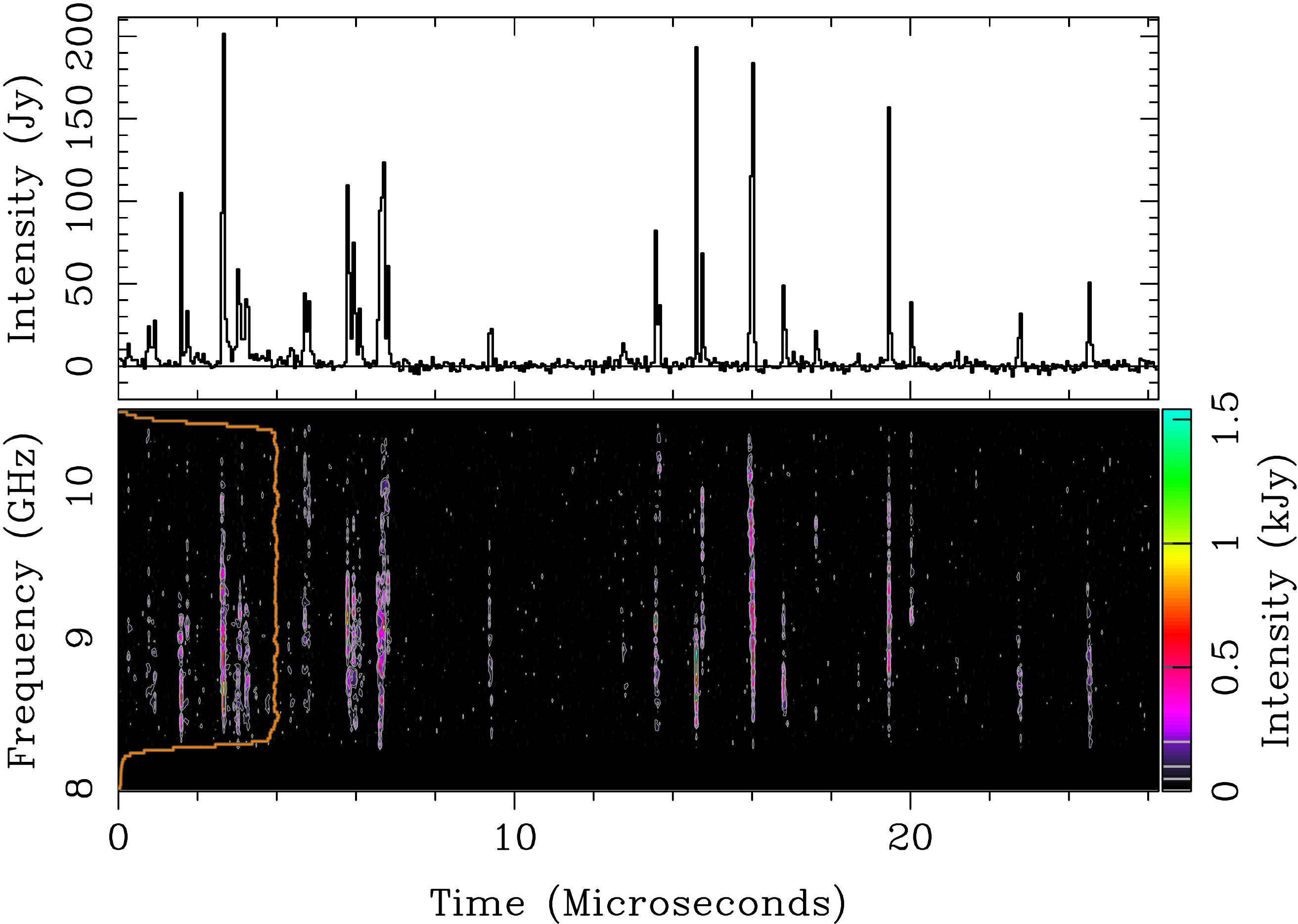}
\includegraphics[width=\columnwidth]{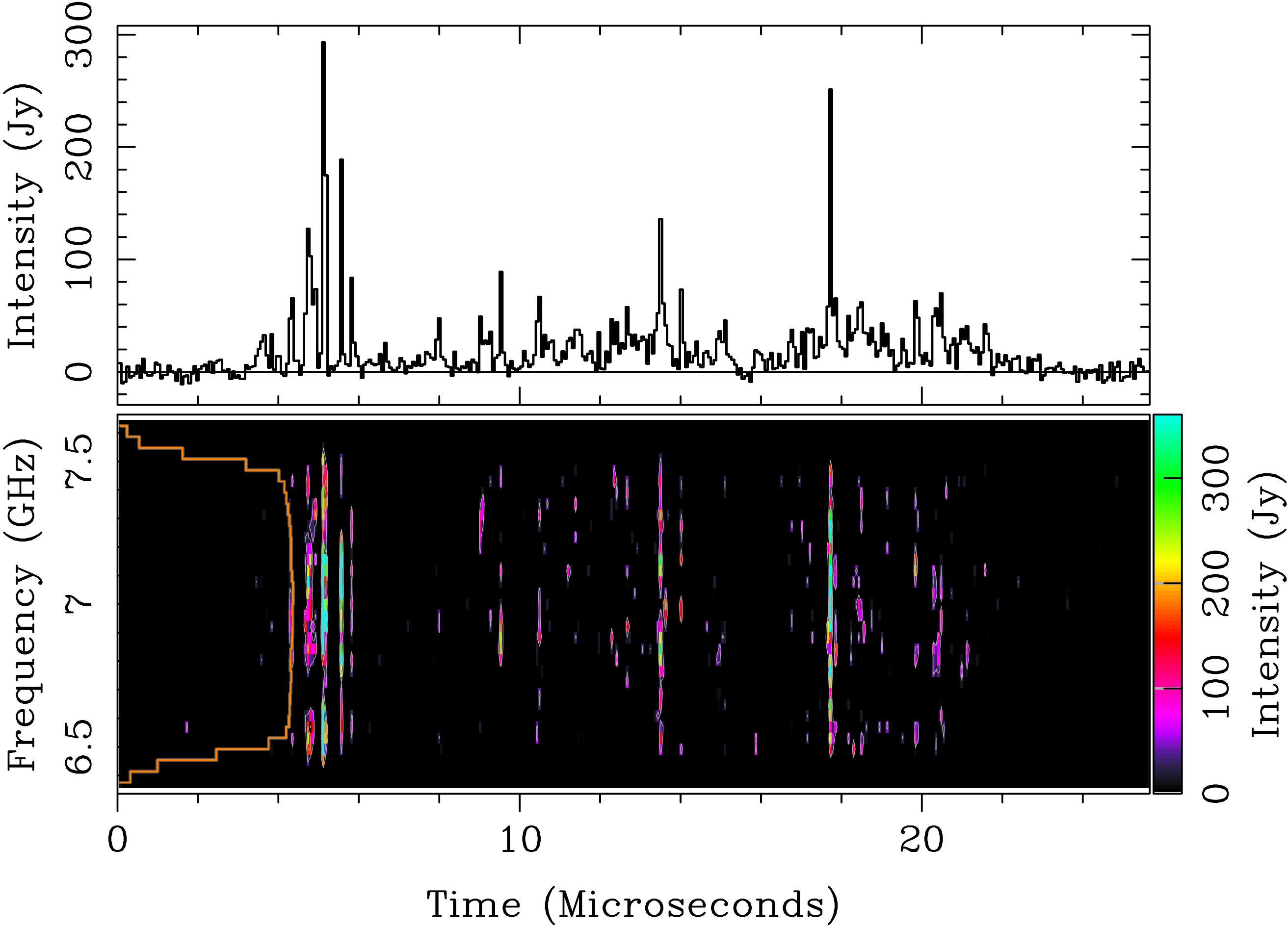}
\caption[]{Two examples of nanoshots. Top: a single Main Pulse between 8 and 10.5 GHz. Shown with time resolution 51.2 ns, spectral resolution 19.5 MHz, and de-dispersed at 56.73736 \pccm3. The contour levels are 0.05, 0.1, 0.2, 0.5 and 1 kJy.  Bottom: a single  Low-Frequency Interpulse seen between 6.4 and 7.6 MHz, shown at time resolution 64.0 ns, spectral resolution 39 MHz, and de-dispersed at DM of 56.74001 \pccm3.  The contour levels are 0.1 and 0.2 kJy. Note, the low-level emission between nanoshots is real, well above the 5.9 Jy level of the off-pulse noise. Both examples illustrate the characteristics of resolved nanoshots: they are extremely short-lived and have frequency spread less than our observed bandwidth. Compare Figure \ref{fig:nanoshot_pol} which shows some of the nanoshots at higher time resolution. }
\label{fig:nanoshots}}
\end{figure}

Very occasionally the nanoshots in a Main Pulse are sufficiently sparse to be studied directly.   Figure \ref{fig:nanoshots} shows such examples, a Main Pulse and a Low-Frequency Interpulse.  Both are shown at relatively low time resolution, to improve the clarity of the dynamic spectra.  \citep[See][for other examples]{Hankins2003,HE2007,EH2016}.  The dynamic spectrum in Figure \ref{fig:nanoshots} shows that well-isolated nanoshots have a relatively narrowband spectrum, with emission bandwidth small compared to our observing bandwidth. Examples are the nanoshots at 9, 22.5 and 24 microseconds in the top panel of Figure \ref{fig:nanoshots}.  The spectrum of other nanoshots extends across our observing band, suggesting they contain substructure which is not resolved at this time resolution. The existence of nanoshots in a Low-Frequency Interpulse is consistent with our other evidence that these two components have similar temporal and spectral characteristics.

\begin{figure}[htb] 
\includegraphics[width=\columnwidth]{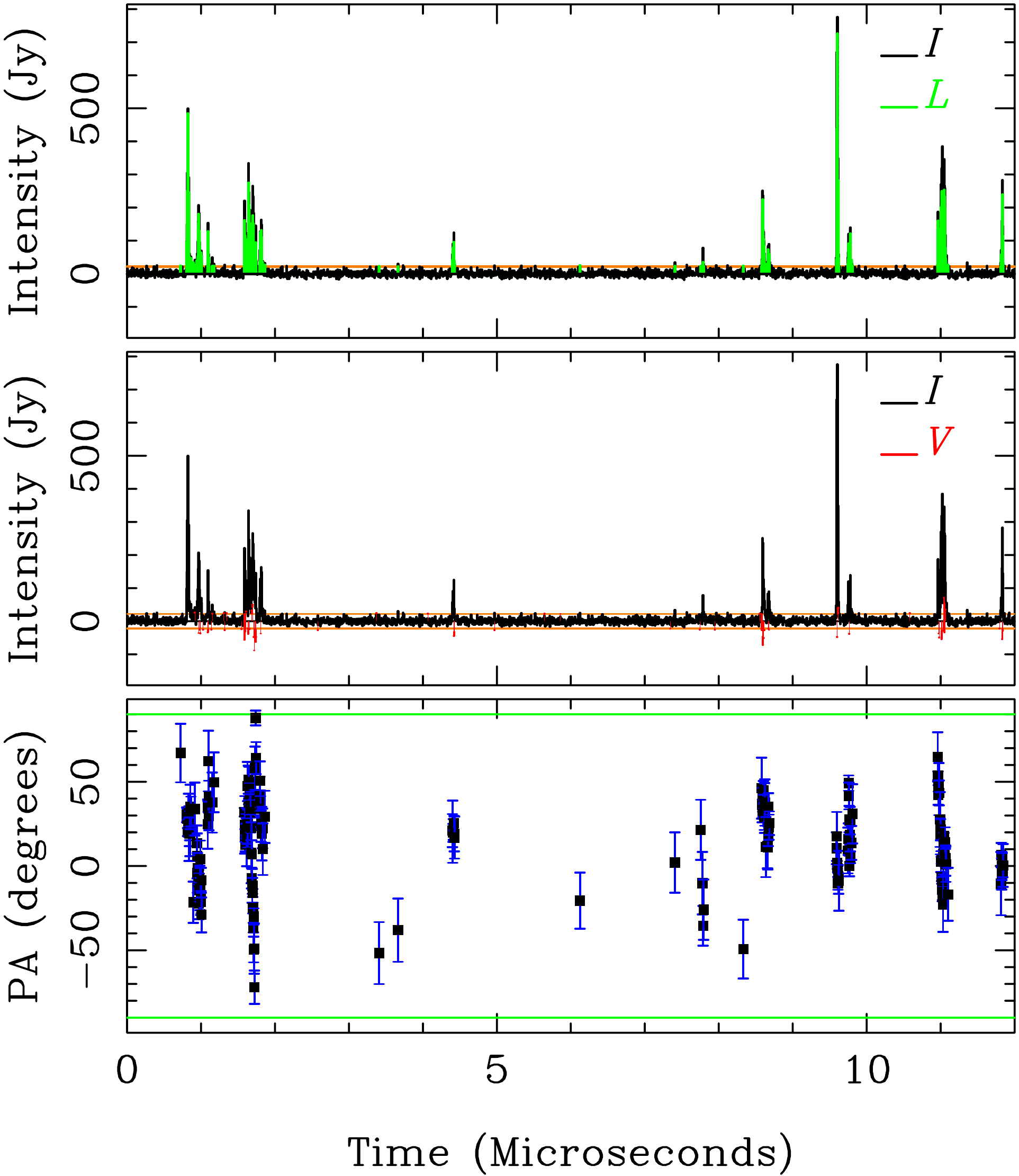}
\caption[]{A zoom into the nanoshot clumps seen between 5 and 16 $\mu$s in upper Figure \ref{fig:nanoshots}, to display their polarization.  Layout is the same as in Figure \ref{fig:MP_ordered_pol}, except that polarization above 3 times the background noise is shown here.   At the displayed time resolution, 4.0 ns, substructure can now be seen in most of the nanoshots, consistent with their blended dynamic spectrum in Figure \ref{fig:nanoshots}.  In this example, the nanoshots are mostly linearly polarized, with some weak circular polarization.  Note that each nanoshot clump has its own position angle behavior.
\label{fig:nanoshot_pol}}
\end{figure}

In Figure \ref{fig:nanoshot_pol} we explore the nanoshot polarization by zooming into a high-time-resolution display of some of the nanoshots in the Main Pulse of Figure \ref{fig:nanoshots}.  At this time resolution, 4.0 ns, substructure can be seen in most of the nanoshots.  The nanoshots in this example are  dominated by linear polarization, although weak circular polarization can also be seen.  Resolved nanoshots we have captured in other pulses show different polarization signatures. For instance, in \citet{EH2016} we show a Main Pulse in which some nanoshots are linearly polarized and others are circularly polarized. 

 Interestingly, each nanoshot in Figure \ref{fig:nanoshot_pol} has its own position angle ``sweep'';  there is no sign of ordered position angle behavior through the entire pulse.  This is also the case for other polarized nanoshots we have captured.  We take this as another indication that the \citet{RC69} model does not apply here.    

\subsection{Emission physics of the nanoshots} 
\label{nanoshot_models}

 We believe nanoshots in the Main Pulse and the Low-Frequency Interpulse provide an important test of radio emission models. A successful model must explain three key characteristics. (1) The basic units of the radio bursts are nanosecond-long ``shots''  of coherent emission.  (2) The product of the center frequency and duration of such a nanoshot obeys  $\nu \delta t \sim O(0.1)$.  (3) The nanoshots are elliptically polarized, with a mix of linear and circular polarization which can vary shot to shot. 
In \citet{EH2016} we survey a range of proposed models.  We argue there that no emission model can explain all three characteristics, but two types of models show the most promise. 

Models based on strong plasma turbulence include soliton collapse \citep{JCW97, JCW98} and self-generated wigglers \citep[e.g.,][]{BBBE88,Sch2003}.  These models naturally produce short-lived bursts of radiation with a timescale related to the inverse of the plasma frequency.   The frequency-duration product can agree with the data (if certain parameters are right).  However, the models as developed so far cannot explain circular polarization. They apply only to extremely strong magnetic fields, which restrict charge motion and lead to linear polarization. 

Models based on the anomalous cyclotron instability in a pair plasma \citep[e.g.,][]{Kazbegi1991,Lyutikov1999} produce circular polarization if the electrons and positrons have different streaming speeds.  However, these models require extreme values of density and magnetic field, which may or may not exist in the upper magnetosphere.  Furthermore, the models as developed so far do not extend to the strong-turbulence regime, so they cannot explain short-lived nanoshots.

 In \citet{EH2016} we suggest a combination of these models might work. We envision a beam-driven instability that creates coherent charge bunches,  in a moderately magnetized plasma which can carry elliptically polarized modes.  However, this idea must remain only speculative, until it can be verified or disproved by future work. 

\section{An unusual Main Pulse at 43 GHz}
\label{section:QbandMainPulse}

\begin{figure}[htb] 
\includegraphics[width=\columnwidth]{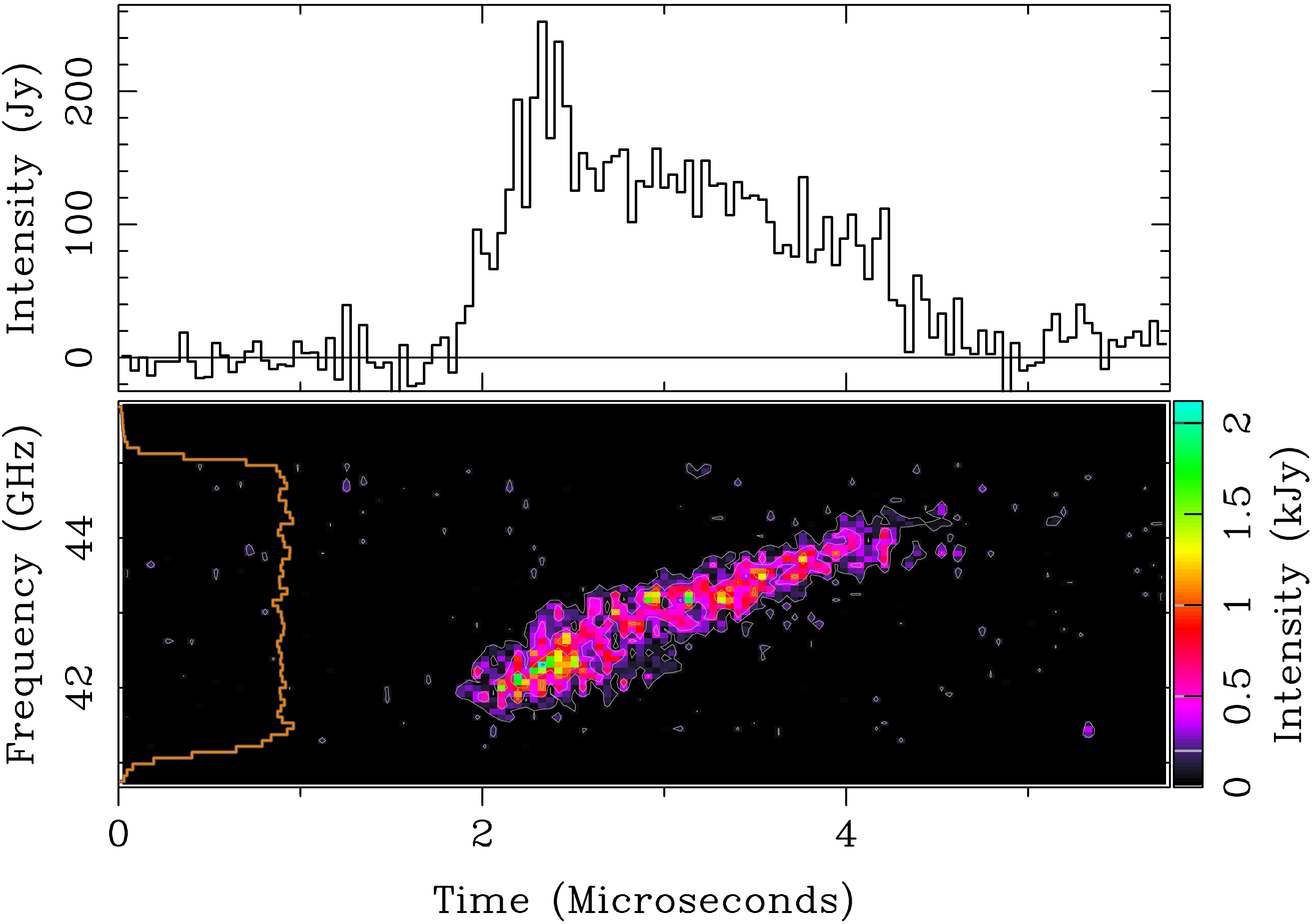}
\caption[]{The total intensity of a Main Pulse recorded at 43.25 GHz and de-dispersed using DM of 56.794 \pccm3 (taken from Jodrell Bank monitoring for our observing date)   is shown with a time resolution of 44.8 ns.  The frequency resolution of the dynamic spectrum is 78 MHz. The Intensity contour levels in the dynamic spectrum are 0.2, 0.5, 1, and 2 kJy.  The off-pulse noise level for the total intensity is 15.2 Jy, and for the dynamic spectrum, 110 Jy.
\label{fig:MP_at_43GHz} }
\end{figure}

Although the Crab pulsar becomes fainter at high frequencies, we managed to capture one Main Pulse at 43 GHz.  This is the highest frequency at which a single pulse has been captured from this pulsar; we spent 12 hours of observing time to catch it.   We show this unusual pulse in Figure \ref{fig:MP_at_43GHz}.  Unlike Main Pulses we have caught at lower frequencies, this pulse lasted 2 $\mu$s, with no sign of internal sub-$\mu$s microbursts.  More strikingly, the upward sweep of the narrow emission band in its  dynamic spectrum is unique among all the Main Pulses we have observed. 

We checked whether the unusual dynamic spectrum of this pulse could be due to dedispersion processing with the wrong DM.   We used the DM determined from Jodrell Bank monthly Crab monitoring at 1.4 GHz appropriate for our observing date.  Dedispersion with a much lower DM ($\sim 41$ \pccm3, well below the $56.794$ \pccm3 Jodrell value) would be necessary to ``straighten up'' the emission spectrum.  We cannot believe that such a ``hole'' in the electron density would have appeared in the Crab Nebula, then vanished again, within only a few days.\footnote{The DM monitored by Jodrell Bank varied by  $\sim 0.20$ pc-cm$^{-3}$ over $\sim 12$ years; our 20 and 23 GHz observations, made only several days earlier, are consistent with the Jodrell Bank value.}  The upward sweep in frequency is also very unlikely to arise from a frequency error in the local oscillator setup or in our processing.  In the dispersion removal parameters, we would have to use a center frequency of 60 GHz to eliminate the frequency sweep. We conclude that the upwards sweep in Figure \ref{fig:MP_at_43GHz} is real -- and completely unexpected.

We also checked whether our identification of this as a Main Pulse could be wrong. Using TEMPO\footnote{see http://tempo.sourceforge.net} to convert the arrival time of this single 43 GHz pulse to rotation phase, we found the pulse arrived within 6 $\mu$s  of the  Main Pulse phase predicted by TEMPO. Furthermore,  although the dynamic spectrum of this pulse is suggestive of a single emission band from a High-Frequency Interpulse, the bandwidth separation $\Delta \nu \simeq 0.06 \nu$ which we measured at lower frequencies \citep[][also Section \ref{section:IP_band_structure}]{HE2007} would put adjacent bands at $\Delta \nu = \pm 2.6$ GHz.  Such bands would easily be visible above or below the single emission band in the dynamic spectrum of Figure \ref{fig:MP_at_43GHz};  their absence is consistent with our identification of this unusual pulse as a Main Pulse. 

Is this 43 GHz Main Pulse an extreme example of dispersion in the pulsar's magnetosphere,  as is common with the High-Frequency Interpulse \citep[][also Section  \ref{section:IP_Dispersion}]{HE2007}? We believe not. If the frequency sweep is due to  dispersion in the magnetosphere, it would require negative dispersion: higher frequencies arriving after lower frequencies.   We know of no astrophysical object where such behavior exists.  

Finally, it is possible that the upward frequency sweep is physically real and due to the emission mechanism itself. With a sample of one, we do not pursue this idea here. 

\section{The High-Frequency Interpulse:  spectral emission bands}
\label{section:IP_band_structure}

In \citet{HE2007} we discovered, to our surprise, that the High-Frequency Interpulse is strikingly different from its Low-Frequency counterpart.  In addition to the $7^{\circ}$ phase shift, the two components have different temporal and spectral characteristics.   High-Frequency Interpulses contain one broad burst of emission, typically several microseconds long, rather than several shorter microbursts.  We have never found sparse nanoshots in a High-Frequency Interpulse. Furthermore, we found a totally unexpected spectrum between 5 and 10 GHz in the High-Frequency Interpulse.   Rather than  continuous, broadband emission, its dynamic spectrum contains a set of discrete \emph{emission bands}.  These bands are not uniformly spaced, as one might expect from harmonic emission.  Instead, we found they are \emph{proportionally spaced},  with frequency separation $\Delta \nu \sim 0.06 \nu$ between 5 and 10 GHz.  In this section we show that the emission bands continue up to at least 30 GHz. 

\begin{figure}[htb] 
\includegraphics[width=\columnwidth]{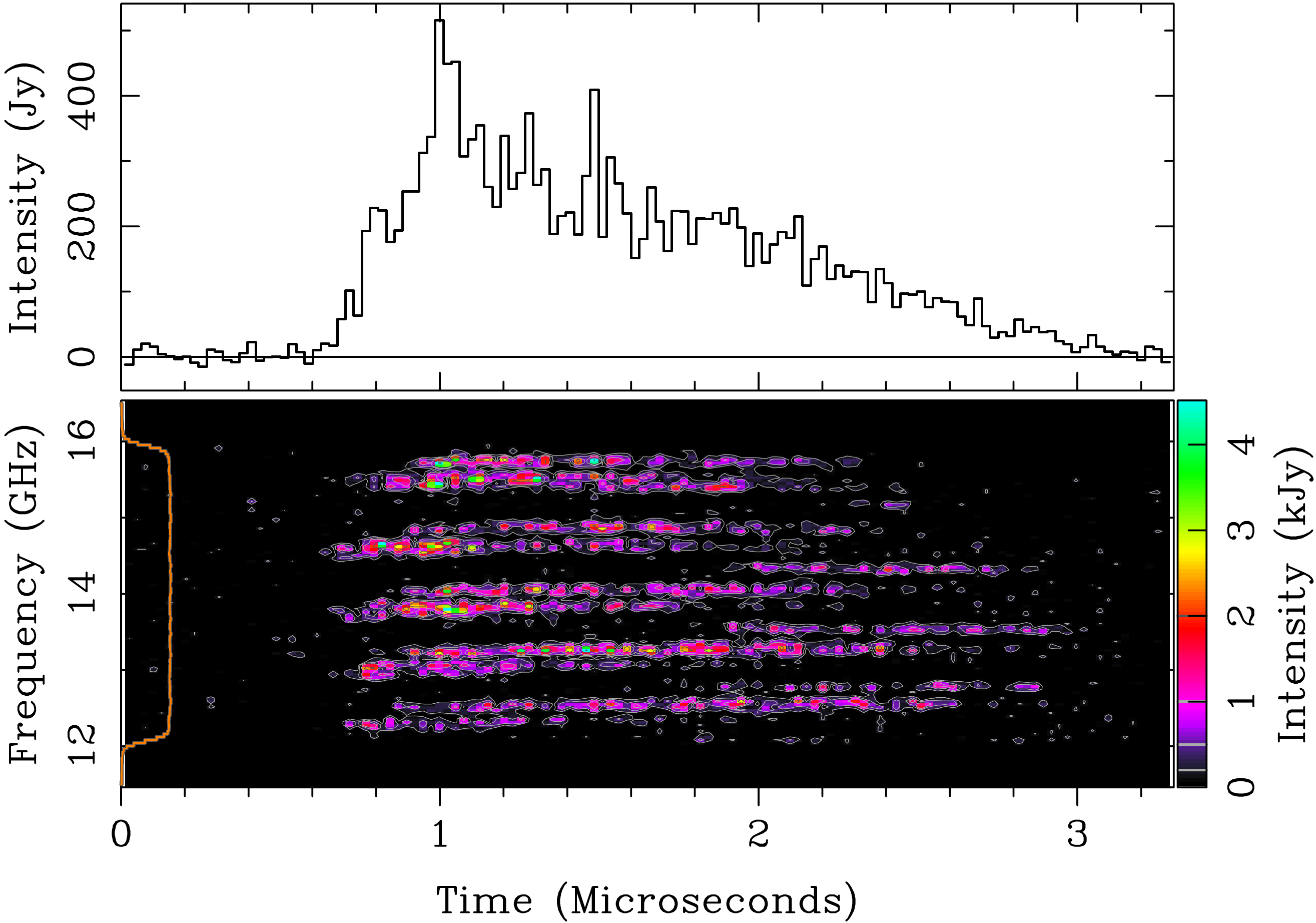}
\includegraphics[width=\columnwidth]{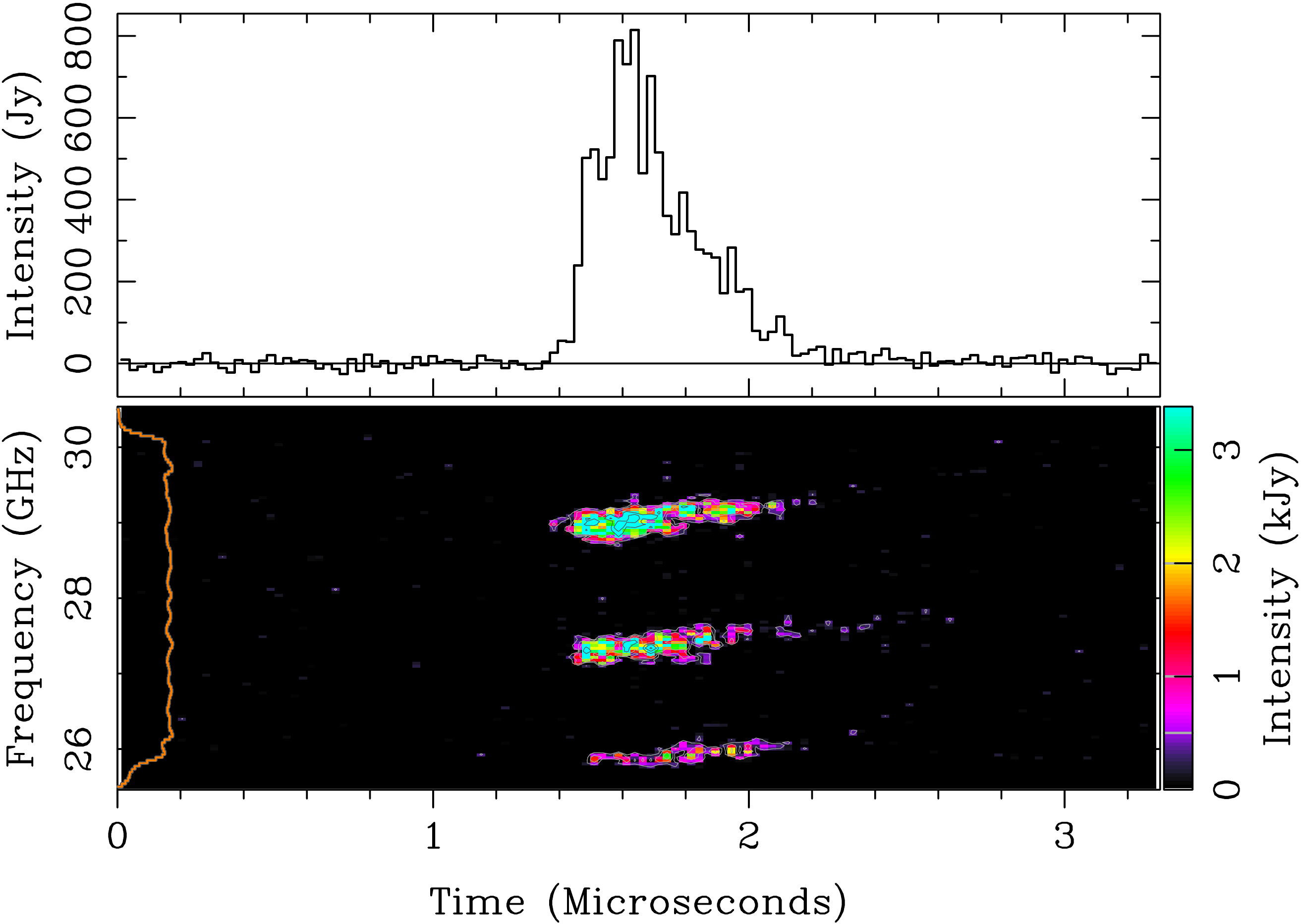}
\caption[]{Two High-Frequency Interpulses observed above 10 GHz.  Top, a pulse captured between 12 and 16 GHz, de-dispersed with DM of 56.801254 \pccm3. The contour levels are 0.2, 0.5, 1, and 2 kJy.  Bottom, a pulse captured between 25.9 and 30.3 GHz, de-dispersed with DM of 56.799770 \pccm3.  The contour levels are 0.5, 1, and 2 kJy. Both pulses are displayed with time resolution 25.6 ns, and spectral resolution 78.125 MHz.  In the top pulse,  five spectral bands can be seen, with three band ``sets'' within each band. In the bottom pulse, three spectral bands can be seen.  Note the larger frequency spacing between bands in the lower example, consistent with the fit in equation \ref{eq:linear_fit}, also shown in Figure \ref{fig:BandSpacing_vs_Frequency}. \label{fig:IP_at_two_freqs}}
\end{figure}

\subsection{Emission bands at higher frequencies}
\label{subsection:FrequencyDependence}

High-Frequency Interpulses are more abundant than Main Pulses above 9 GHz \citep{Hankins2015}; we were able to catch 240 new examples between 12 and 30 GHz. Figure \ref{fig:IP_at_two_freqs} shows two typical examples, at 14 and 28 GHz.  These new data show that the character of the High-Frequency Interpulse continues unchanged at these high frequencies. The spectral bands are still there, and their spacing increases with frequency, as expected from our previous results. 

\begin{figure}[htb] 
\includegraphics[width=\columnwidth]{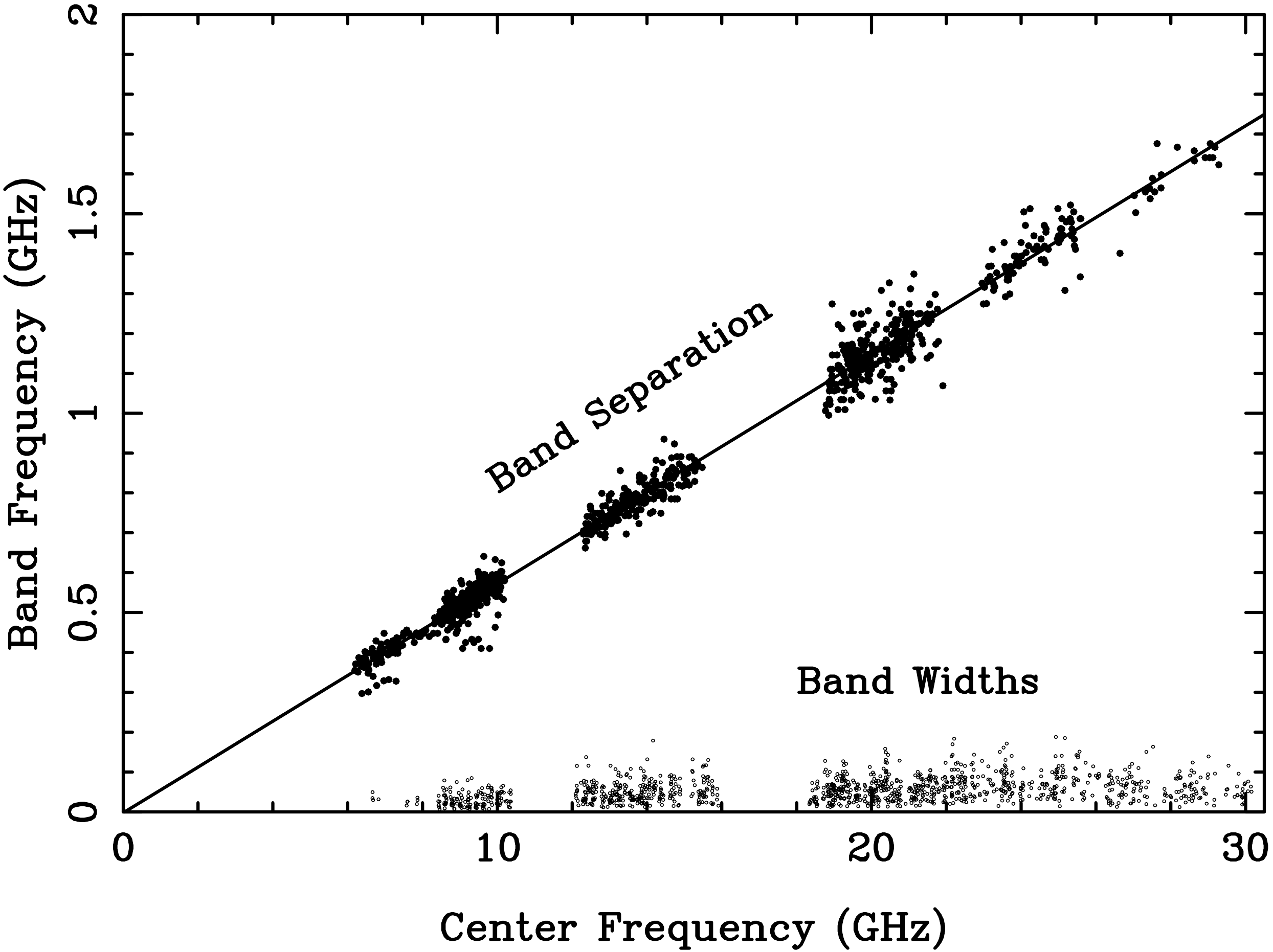}
\caption[]{The Interpulse emission band spacing is plotted against the band center frequency, using data from this paper and from \citet{HE2007}.  The plot is extended down to (0,0) to emphasize that the first term of the linear fit is consistent  with zero. The band widths, measured by Gaussian fits,  are shown on the same scale: the bands are much narrower than their spacing. 
\label{fig:BandSpacing_vs_Frequency} }
\end{figure}

To explore the band spacing at these new frequencies,  we measured the band center frequencies of many of the High-Frequency Interpulses we captured above 10 GHz. In Figure \ref{fig:BandSpacing_vs_Frequency} we combine these results with our earlier measurements \citep{HE2007}.  It is clear that our previous linear relation between band spacing ($\Delta \nu$) and band-center frequency ($\nu$) continues to 30 GHz.  Specifically, we fit the band spacings with a linear function,
\begin{equation}
\Delta \nu = -0.0023\pm0.0030 +(0.0574\pm0.0002)\nu   
\label{eq:linear_fit} 
\end{equation}
finding a reduced chi-squared value of 1.3949 (all frequencies measured in GHz).  We also tested quadratic fits. We found no evidence of a quadratic relationship between $\Delta \nu$ and $\nu$, nor did we find any evidence of an offset from the origin. Our band spacing measurements are are consistent with a straight line passing through the origin and having a slope of about 6\%.  

Although  the data in Figure \ref{fig:BandSpacing_vs_Frequency} extend from 5 to 30 GHz, we have no direct evidence that the bands themselves exist throughout that range in any one pulse. Our largest bandwidth is 4 GHz, too narrow to sample more than a few of the bands at any one time.  However, we have never seen  a High-Frequency Interpulse that did  not show emission bands throughout our observing bandwidth.  This fact, plus the clear linear relation in Figure \ref{fig:BandSpacing_vs_Frequency}, suggests to us that every High-Frequency Interpulse contains continuous spectral bands from (at least) 5 to 30 GHz. If this is the case, there must be about 30 such bands over that full range. 

We also measured the spectral widths of the Interpulse bands, by fitting Gaussians to individual bands. Although some of the fits were contaminated or biased because of overlapping band sets, the general behavior of the bandwidths can be seen in Figure \ref{fig:BandSpacing_vs_Frequency}.  We find that the spectral width of an emission band does {\it not} increase as fast as the band separation does. 

We occasionally see emission bands in the High-Frequency Interpulse that appear to drift upwards in frequency, such as in the lower panel of Figure \ref{fig:IP_at_two_freqs}. However, upon close examination of many pulses, we find the appearance of upward drift is usually caused by the superposition of new band sets that begin slightly later in time at slightly higher band frequencies.  This can be seen in the upper panel of Figure \ref{fig:IP_at_two_freqs};  another example is shown in Figure 7 of \citet{HE2007}.   We have never seen any emission bands in a High-Frequency Interpulse with the strong upward frequency drift -- about 5 percent of the band center -- that we found in the pulse shown in Figure \ref{fig:MP_at_43GHz}.
 
\subsection{Is there frequency memory in the emission bands?}

Searching for more clues on the origin of the emission bands, we also explored whether bands in separate High-Frequency Interpulses have any ``frequency memory.'' Are the band frequencies, as well as their fractional spacing, steady over time?

\begin{figure}[htb] 
\includegraphics[width=\columnwidth]{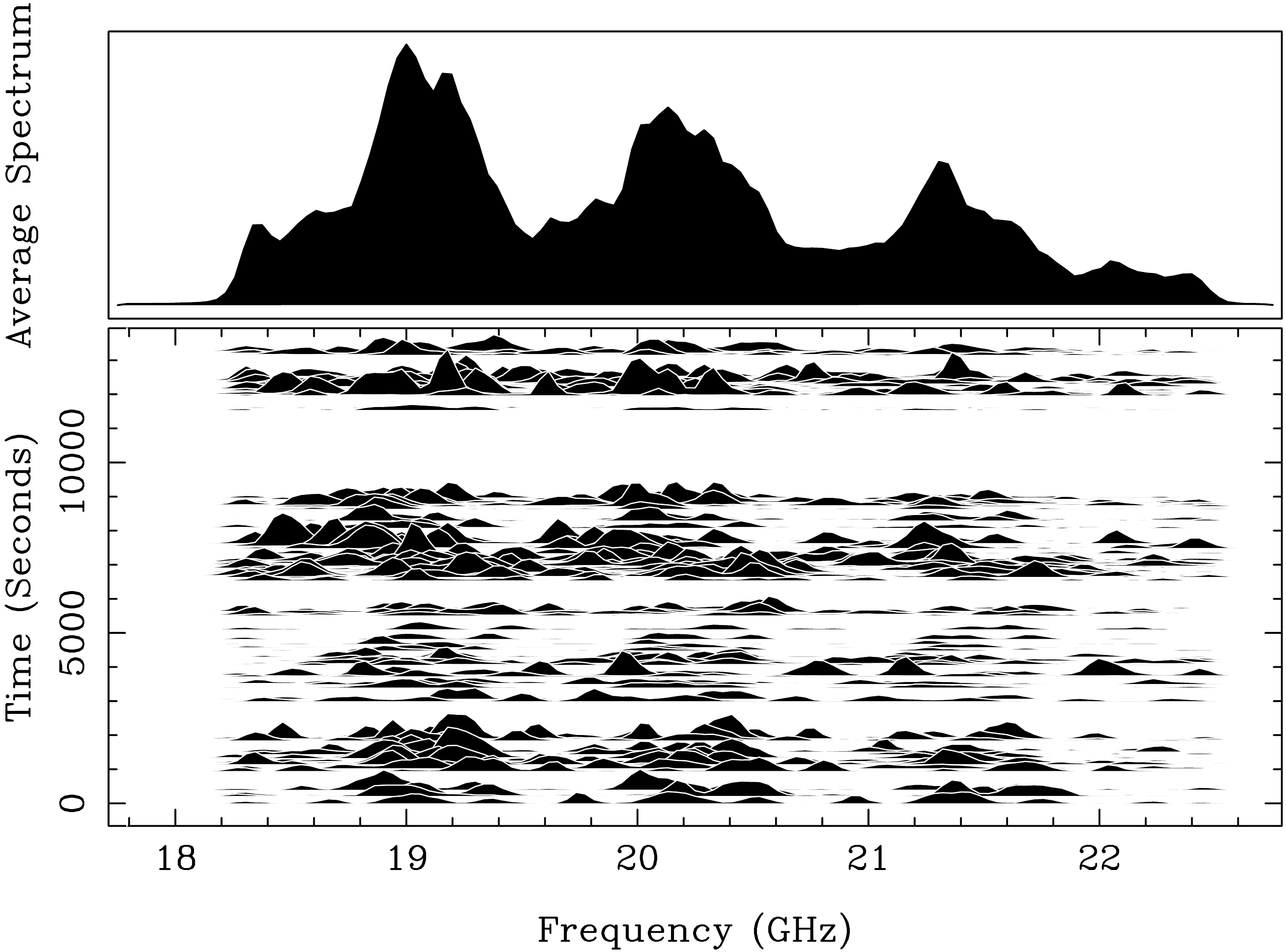}
\includegraphics[width=\columnwidth]{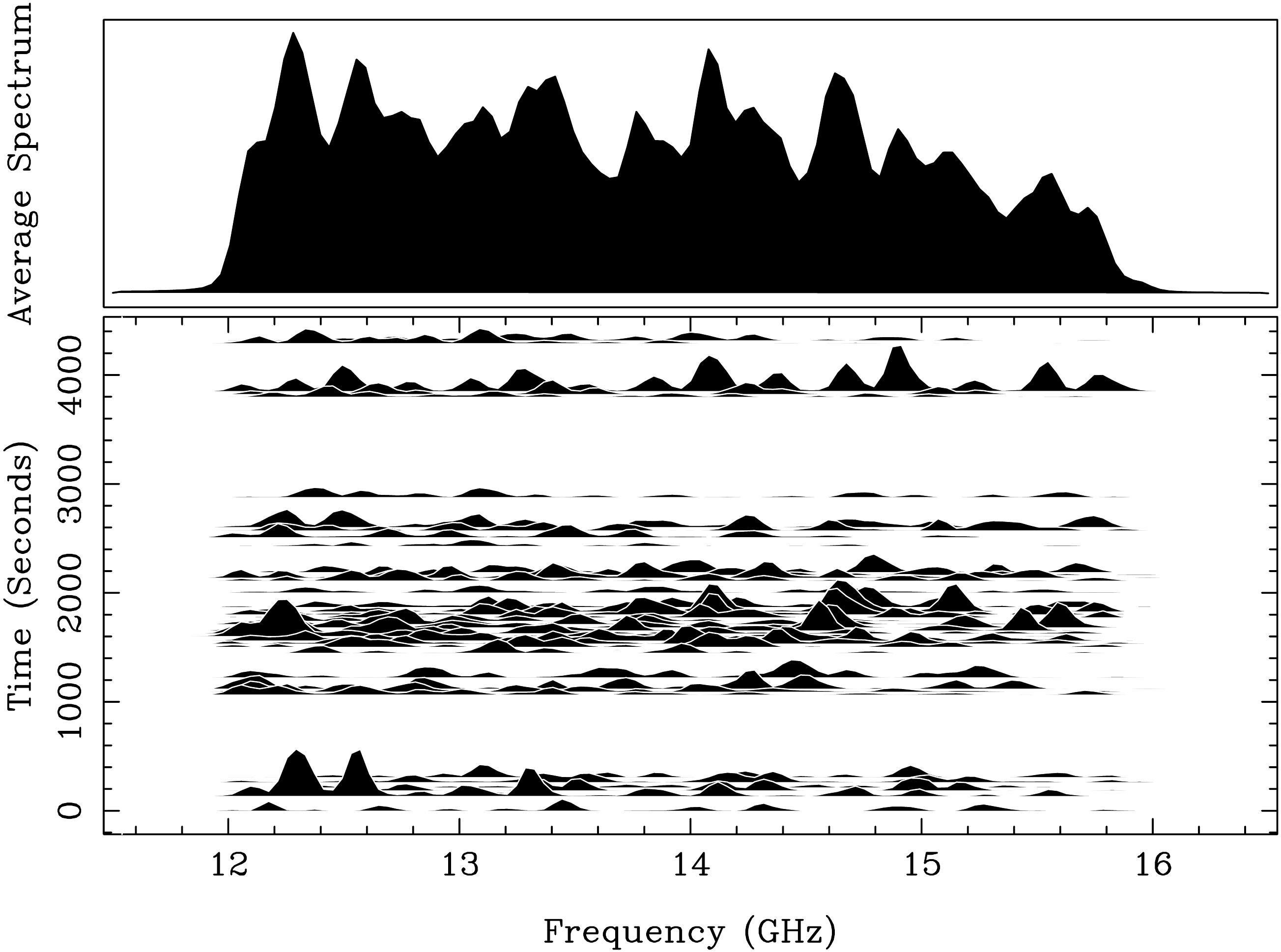}
\caption[]{Two examples of individual dynamic spectra for all pulses recorded during one observing day, with the average spectrum for the day shown in the top panel.  The example on the top appears to show preferred frequencies over the entire day;  the example on the bottom does not. In each example, gaps in the data are due to system, pointing and calibration checks of the telescope.}
\label{fig:GrandAvSpec}
\end{figure}

One way to check this uses multiple pulses within one rotation period. We know the phase windows associated with components of the mean profile are much wider than the duration of a single pulse \citep{Hankins2015}.  For instance,  at $\gtw 5$ GHz, individual High-Frequency Interpulses last only  $\sim 2~ \mu$s, much less than the $\sim 720 ~\mu$s width of the associated component of the mean profile. Our sampling window is several times wider than the component width, and we find that about 1/4 of the data records show multiple High-Frequency Interpulses separated up to several hundred microseconds. If the bands have frequency memory,  we would expect each of the multiple High-Frequency Interpulses in one rotation period to have similar band frequencies.  However,  when we inspected examples of multiple pulses, we did \emph{not} find any consistent frequency memory.

We also checked whether the emission bands within the Interpulse have preferred frequencies within a few-hour observing run.  Here the result was mixed. Figure \ref{fig:GrandAvSpec} shows two examples from days where a sufficient number of pulses was recorded. In the upper panel it is clear that both the single-pulse and daily average spectra show favored band frequencies at about 19, 20.2 and 21.4 GHz. The lower panel shows what appears to be a random set of single-pulse band frequencies with no preferred frequencies in the day average spectrum.  Overall, we conclude there is no strong evidence for band frequency memory.     

\subsection{Emission physics for the High-Frequency Interpulse}

Neither of the models we discussed in Section \ref{nanoshot_models} can account for the emission bands in the High-Frequency Interpulse. This is also true of the other usual suspects for pulsar radio emission \citep[e.g.,][]{Melrose1995,EH2016}.  However, since the bands were discovered,  several {\it ad hoc} models have  been suggested to explain the spectral bands. 
\subsubsection{Direct emission models}

{\it Resonant cyclotron emission.} \citet{Lyutikov2007} suggested the bands are harmonic emission of plasma waves excited by the anomalous cyclotron instability. With his specific parameter choices, which require high plasma densities close to the light cylinder,  the spacing of the first few harmonics approximately agrees with the 6--10 GHz results we presented in \citet{HE2007}.  Unfortunately,  higher-order  harmonics of the the wave mode he proposes do not obey the simple 6\% proportionality we observe (even if 30 harmonics could somehow be excited in the first place, which may be unlikely;  M. Lyutikov, private communication 2015).

{\it Beamed superluminal emission.} \citet{Ardavan1994} proposed that beamed emission from a superluminal polarization current pattern, outside the light cylinder, accounts for pulsar radio emission.  \citet{Ardavan2008} applied this to the Crab pulsar, suggesting the spectral bands are due to monochromatic oscillations of the  polarization current. With their parameter choices, the first few bands they predict do approximately match our 6--10 GHz data in \citet{HE2007}.  However, their model also predicts band spacing increasing with frequency, as $\Delta \nu \propto \nu^{3/2}$.  This disagrees with the simple $\Delta \nu \simeq 0.06 \nu$  proportionality we observe.

{\it Double plasma resonance.} The emission bands we find in the High-Frequency Interpulse are reminiscent of so-called {\it zebra bands} found in type IV solar flares \citep[e.g.,][]{Chernov2005}.  The most successful model for zebra bands is a double plasma resonance,  which requires the upper hybrid frequency be an integer multiple of the cyclotron frequency \citep[e.g.,][]{Chen11,EH2016}.  If the spatial structure of the local density and magnetic field allows this resonance to be satisfied at more than one location, a set of double-resonant harmonics can be excited.  \citet{Zhel2012} suggested this model may also explain the High-Frequency Interpulse.  They note that low magnetic fields and high plasma density are needed, and speculate that such conditions may exist close to the light cylinder.  They did not, however, discuss the specific field and density structures that would be needed to create the uniformly proportional spacing we observe.

\subsubsection{Propagation models}

{\it Stimulated Compton Scattering.} Petrova has suggested that the High-Frequency Interpulse results from stimulated Compton scattering of the Main Pulse.  In \citet{Pet2008} she argues that induced Compton scattering of the Main Pulse creates the High-Frequency Interpulse.  Alternatively, in \citet{Pet2009} she proposes that Compton scattering of the Main Pulse creates the Precursor, and the Precursor itself is then scattered to create the High-Frequency Interpulse.  Both of these models are challenged by the emission bands we observe in the High-Frequency Interpulse.  \citet{Pet2008} suggests that scattering of individual nanoshots in the Main Pulse, which concentrates their power toward the edges of the shot's emission band, can lead to bands in the dynamic spectra. However, she presents no explanation of how this model can lead to the regular, proportional spacing we observe over a factor of six in frequency.

{\it Interference models.} In \citet{HE2007} we speculated that the emission bands may be a propagation effect.  If a radiation beam could be split coherently, perhaps by reflection, it may interfere with itself.  Alternatively, if low-density cavities exist in the plasma, they might impose a discrete frequency structure on escaping radiation. Similar models have been proposed for zebra bands \citep[e.g.,][]{Led2001,Lab2003}. We also noted that the incoming radiation must be broadband -- extending at least from 5 to 30 GHz -- in order for these models to work.  This requirement appears to exclude many standard pulsar radiation mechanisms which peak around the local plasma frequency.  We suggested that linear acceleration emission, for instance in a double layers within gap regions \citep[e.g.,][]{K1990}, may be a possible alternative.  Unfortunately, although we find such models attractive, we have not come up with a way to make the small, long-lived  plasma structures in the magnetosphere that such models require.

\section{Polarization of the High-Frequency Interpulse}
\label{section:IP_Polarization}

Observations by \citet{MH1999} showed that the High-Frequency Interpulse is strongly polarized in mean profiles.  At 5 and 8 GHz, they found that component shows 50--70\% linear polarization.  \cite{Jessner2010} found this also true for a few individual High-Frequency Interpulses at 15 GHz. 

\subsection{Data: uniform polarization position angle}

\begin{figure}[htb] 
\includegraphics[width=\columnwidth]{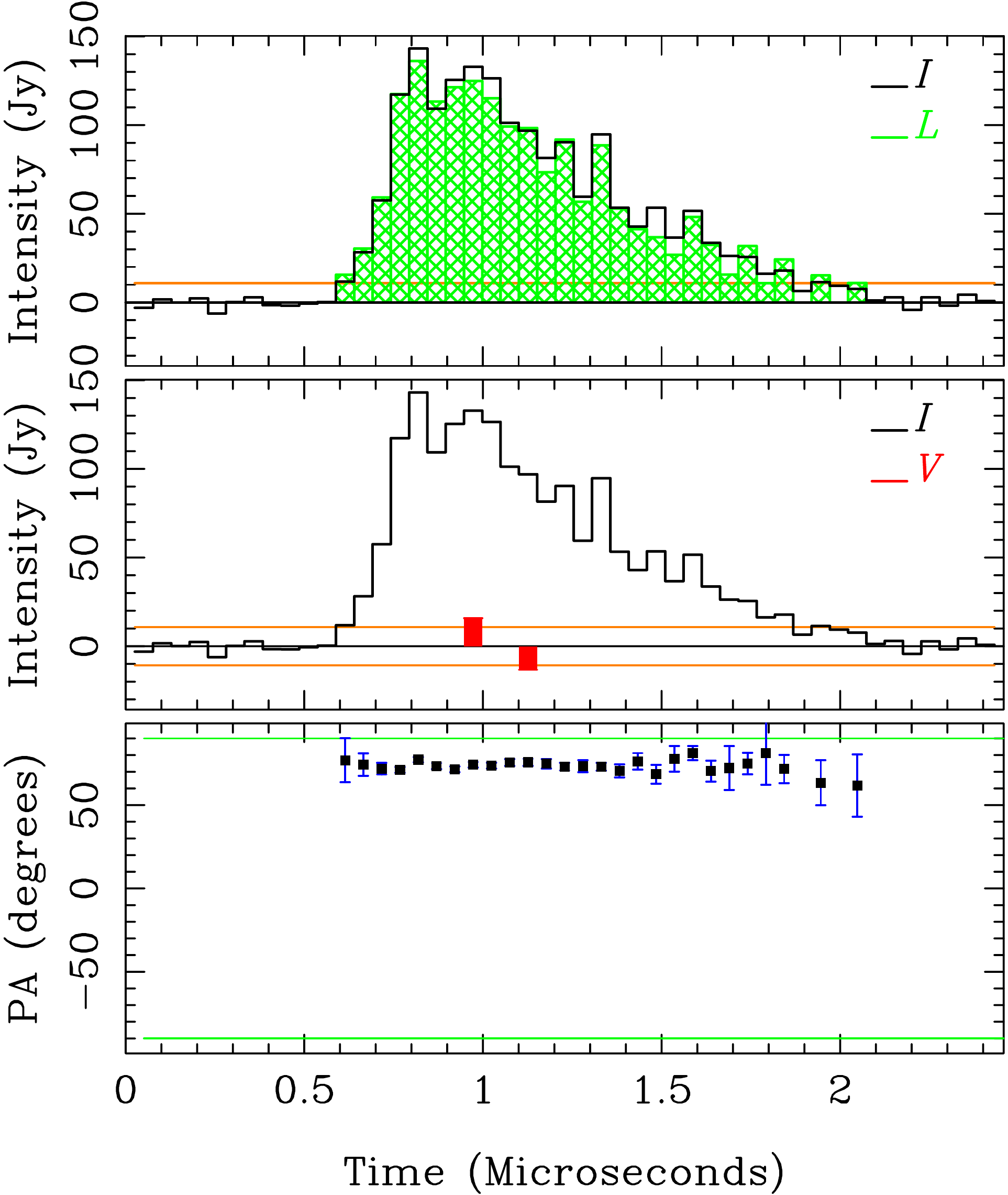}
\caption{Polarization of a typical High-Frequency Interpulse, captured between 17.8 and 22.8 GHz, de-dispersed at DM of 56.79476 \pccm3, and displayed at 51.2 ns time resolution.  Layout the same as in Figure \ref{fig:MP_ordered_pol}, except that polarization greater than 3 times off-pulse rms noise level is shown.  This example is typical of the majority of High-Frequency Interpulses we have observed:  it shows strong linear polarization, only weak circular polarization, and a nearly constant position angle across the pulse. }
\label{fig:HFIP_flatPA}
\end{figure}

Our data support these results. High-Frequency Interpulses that we have captured between 5 and 24 GHz are similarly polarized, typically with 80--90\%  linear polarization.  Circular polarization is occasionally seen, but is generally weak (10--20\%) or undetectable. Figure \ref{fig:HFIP_flatPA} shows one such example; in \citet{EH2016} we show two other examples.  The polarization position angle in these examples remains nearly constant throughout the pulse, in agreement with the results of \citet{Jessner2010} for High-Frequency Interpulses at 15 GHz.

\begin{figure}[htb] 
\includegraphics[width=\columnwidth]{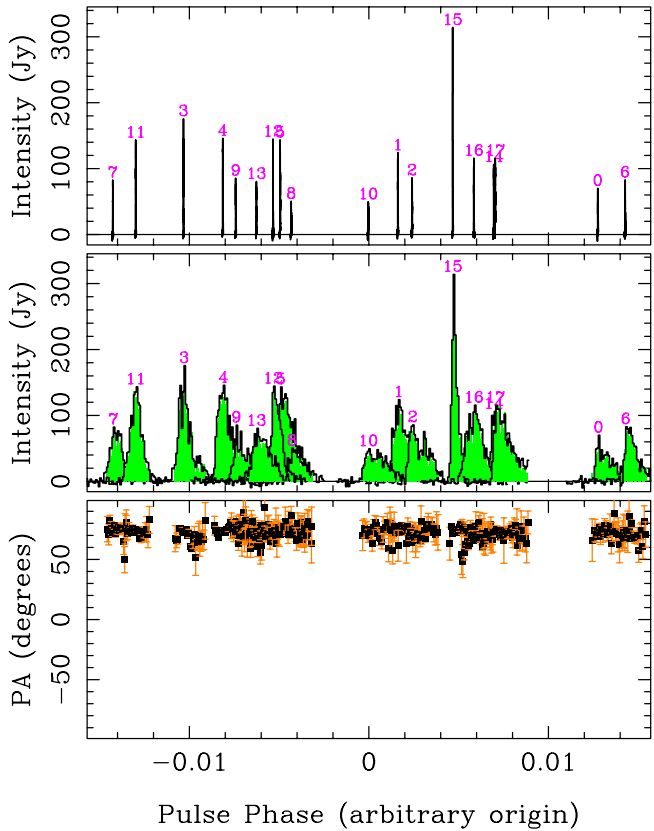}
\caption[]{Polarization of a sequence of High-Frequency Interpulses, captured within 13 minutes on one observing run and displayed with 51.2 ns time resolution.  Upper panel shows the pulses plotted as a function of the rotation phase at which they appeared (a 0.03 span of phase corresponds to $\sim 1$ msec of time).  The numbers give the order in which the pulses were captured, but do not represent sequential rotations of the star. The middle panel shows the same information, but now the time scale of each pulse is expanded by a factor of 50 to show structural details. The lower panel shows the polarization position angle,  also shown expanded in pulse phase by a factor of 50. The position angles of each pulse remain approximately constant, independent of the rotation phase at which the pulse arrived.  Processing is the same as in Figure \ref{fig:HFIP_flatPA};  the example pulse in that figure is number 5, at $-0.005$ pulse phase, in this figure.}
\label{fig:HFIP_ppr_set}
\end{figure}

Not only do we see no significant position angle evolution within most individual High-Frequency Interpulses,  but the position angle is also independent of rotation phase at which the pulse occurred.  Figure  \ref{fig:HFIP_ppr_set} illustrates this for 18 pulses captured in one observing day:  the position angle remains approximately constant across the full range of rotation phase at which we captured pulses.  The upper panel of this Figure shows both the pulses and their arrival phase on the same timescale.  Because individual pulses are very short-lived compared to the 0.03 window of rotation phase within which the pulses arrived on this observing day,  no detail can be seen within any of the pulses in this display.  Therefore, in the two lower panels of this Figure we expand the timescale of each pulse by factor of 50, while maintaining their relative arrival phases,  so that structural details can be seen.  

\begin{figure}[htb] 
\includegraphics[width=\columnwidth]{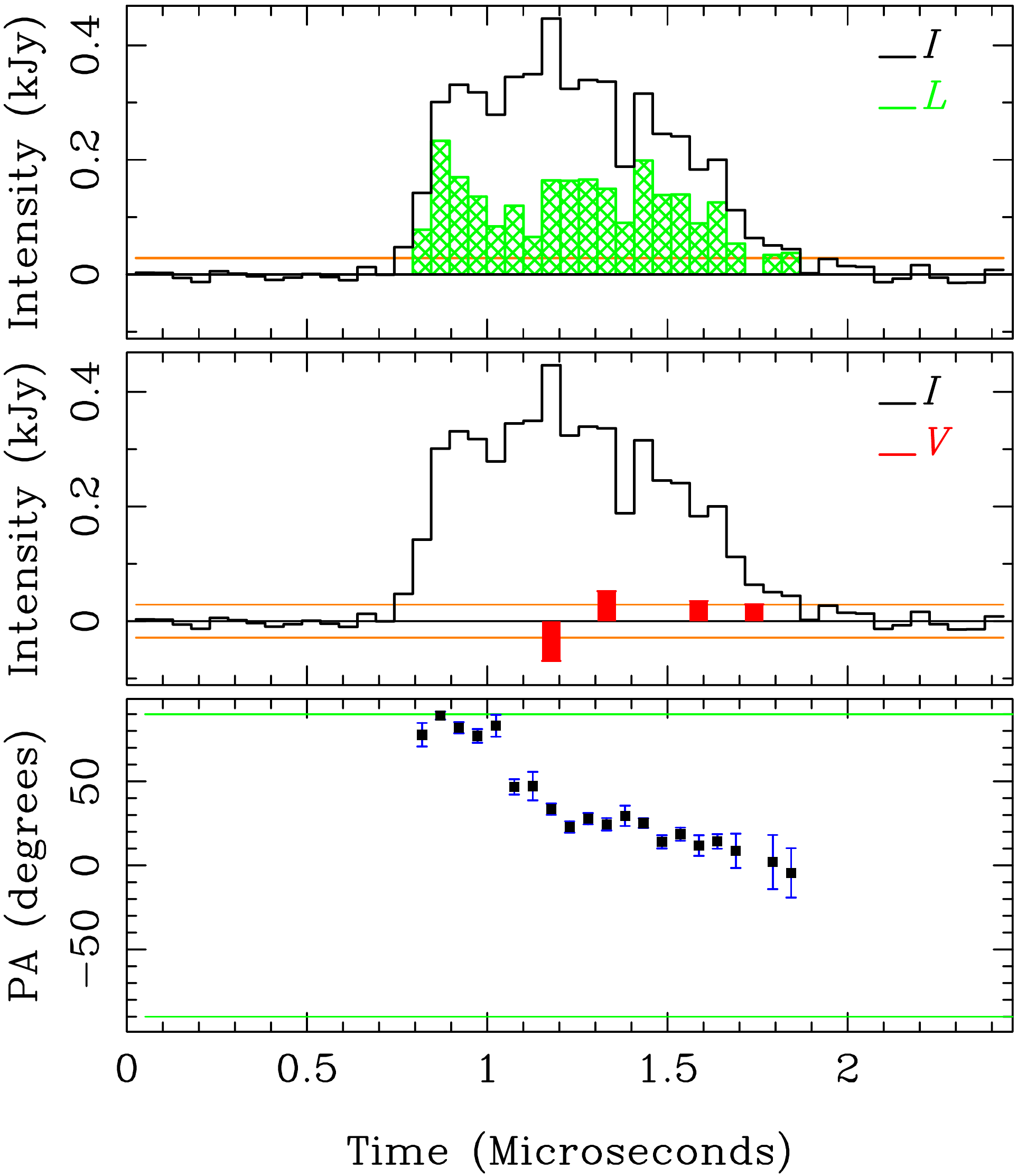}
\caption[]{Polarization of a High-Frequency Interpulse which shows significant position angle rotation.   Captured between 17.7 and 22.4 GHz, de-dispersed at DM of 56.80600, and displayed at 51.2 ns time resolution.  Layout the same as in Figure \ref{fig:MP_ordered_pol}, except that polarization greater than 3 times off-pulse rms noise level is shown.  The linear polarization in this pulse is weaker than that in Figure \ref{fig:HFIP_flatPA}, which is typical of High-Frequency Interpulses which show position angle rotation.}
\label{fig:HFIP_nonflatPA}
\end{figure}

Occasionally, however, we captured some High-Frequency Interpulses with significant position angle rotation across the pulse.  Figure \ref{fig:HFIP_nonflatPA} shows one such example.  Such pulses often have weaker fractional polarization than pulses with
constant position angles.  

\subsection{Constraints on the emission region}

Although we do not know the magnetospheric location of the High-Frequency Interpulse emission region, we can make two general statements based on our polarization data.  (1) The emission region must be spatially localized.  If it were extended along the full caustics, thought to be the origin of pulsed high-energy emission, the High-Frequency Interpulse would be significantly depolarized \citep[e.g.][]{Dyks04}, which is not the case.  (2) The magnetic field direction must be approximately constant within the emission region, and must remain stable during the duration of each dayÕs observing run.  Significant variations in the field direction would be reflected in variable polarization position angles for single High-Frequency Interposes, which is not the case.

\section{Dispersion of the High-Frequency Interpulse} 
\label{section:IP_Dispersion}

The 9 GHz example pulses we presented in \citet{HE2007} showed that single High-Frequency Interpulses can have higher dispersion measure than single Main Pulses.  In this section we explore this result in our full data set.

\subsection{Data:  excess dispersion measure}
\label{DM_analysis_details}

Because inspecting individual pulses (as in Section \ref{DM_note}) is daunting for large numbers of pulses, we used the following methods to analyze our full data set. We first de-dispersed each pulse using the monthly DM values given by Jodrell Bank  (``JB''; Section \ref{DM_note}), spline interpolated to each observing epoch.

We split the Fourier Transform of each dedispersed pulse into the top and bottom halves of our observing band, then transformed each half band back to the time domain and obtained the unsmoothed intensity. We then carried out a series of correlations to determine the time delay between the two band halves. We formed the cross-correlation function (CCF) of the upper and lower bands, and found the time at which that CCF peaked by fitting a parabola around the CCF maximum. We call this ``method 1.''   We also formed the autocorrelation functions (ACFs) of the intensity profiles in the  upper and lower bands, and took their geometric mean. To find the CCF peak we then cross correlated the CCF and the mean ACF, and found the time lag at which that cross-correlation peaked.  We call this ``method 2.'' In Figures \ref{fig:IP_dispersion_singles} and \ref{fig:AveDMHist} we use method 2, which reduces the noise in the final correlation.  In Figure \ref{fig:DMbyBand} we compare both methods and verify that they give similar results.

Once the time delay between the upper and lower band halves is determined, we convert it to an excess DM for each pulse, relative to the JB value: $\DM_{\rm tot} = \DM_{\rm JB}$ + \DDM. Here we use the usual relation between pulse arrival time, $t_{\rm p}$, and center frequency, $\nu$:
\begin{equation}
A \nu^2 t_{\rm p}(\nu) =  \DM_{\rm tot} 
\label{tp_DM}
\end{equation}
where the constant $A = 2.41 \times 10^{-16}$ s-pc-cm$^{-3}$, and $\DM \equiv \int n_{\rm e} dL$, measured in \pccm3, is defined as the integral of the election density $n_{\rm e}$ over the path length $L$. 

We note that this expression for $t_{\rm p}(\nu)$ invokes the $\nu^{-2}$ functional form which holds for cold plasma, such as the interstellar medium. There is no {\it a priori} reason to assume this dispersion law also holds for the magnetospheric plasma. However, because the magnetospheric dispersion law is unknown, we cannot specify a better form. We might expect a different frequency dependence in the magnetosphere to appear as a frequency dependence in our \DDM\ calculations, but in practice that turned out not to be the case.

\subsubsection{Example:  one illustrative day}

\begin{figure}[htb] 
\includegraphics[width=\columnwidth]{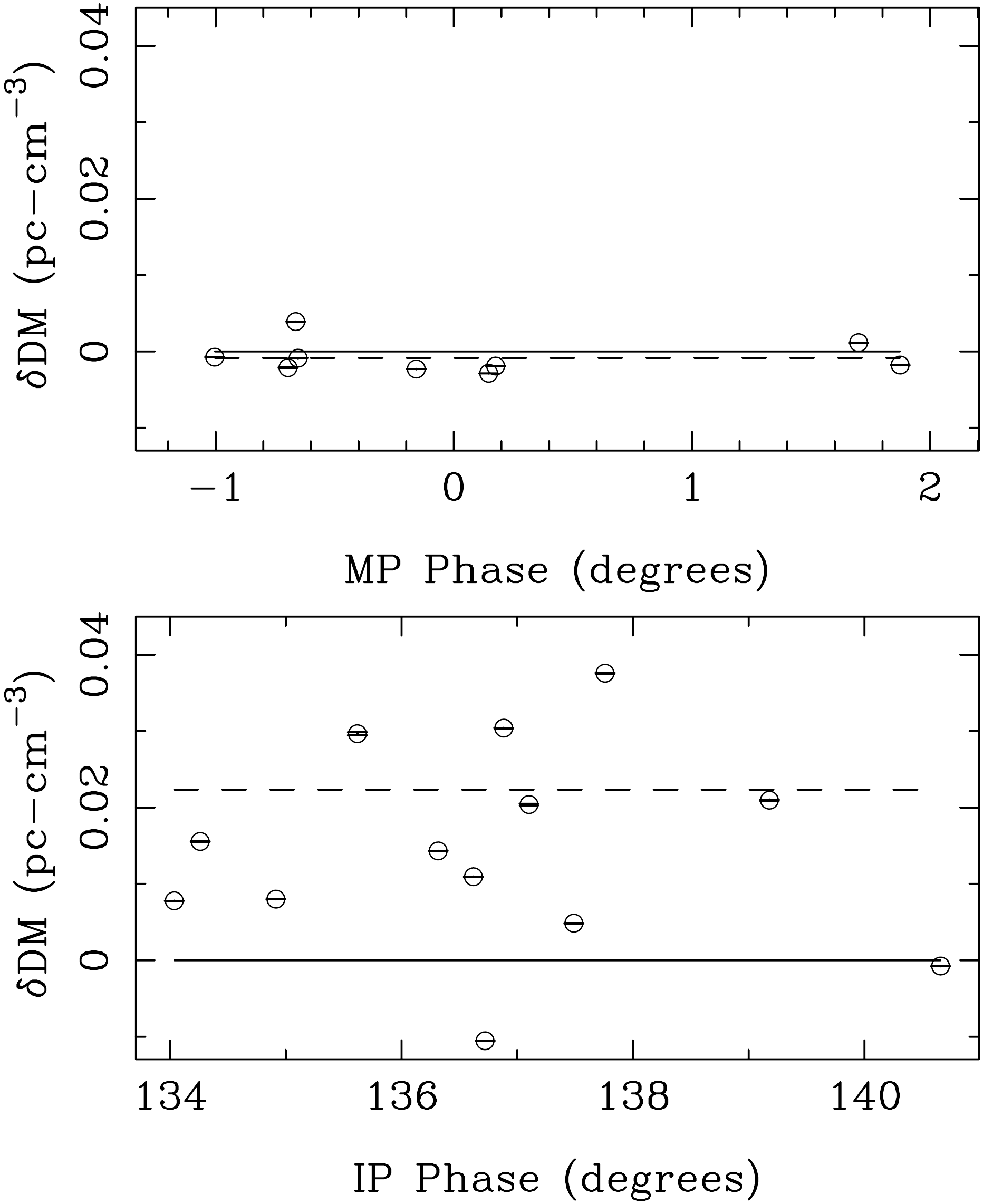}
\caption[]{Excess dispersion measures, \DDM, for individual Main Pulses (top) and High-Frequency Interpulses (bottom) captured between 8.0 and 10.5 GHz, within 100 minutes on one observing day, given relative to the Jodrell Bank value.  Dotted lines show the mean \DDM\ value for each pulse type; solid lines correspond to the JB value (\DDM $\ = 0$).  The \DDM\ values for the  Main Pulses are approximately consistent with the JB value, but the \DDM\ values for the High-Frequency Interpulses are generally larger than the JB value.  The scatter in the High-Frequency Interpulse DM values is real.  The formal error bars, showing the uncertainty of the peak of a parabolic fit to the  maximum of the CCF -- are smaller than the plotted points.  Outliers with $|\delta$(DM)$| > 0.02$ \pccm3 for the Main Pulse, $|\delta$(DM)$| > 0.04$ \pccm3 for the High-Frequency Interpulse,  or fit uncertainty $\sigma_{\rm DM} > 0.01$ \pccm3 for either pulse have been deleted.  These outliers result from unusual frequency-dependent differences between the pulse shapes in the two bandpass halves. \label{fig:IP_dispersion_singles}
}
\end{figure}

To illustrate the excess dispersion measures we find for single pulses, in Figure \ref{fig:IP_dispersion_singles} we show the \DDM\ values for one observing day which contained a good number of both Main Pulses and High-Frequency Interpulses.  We see two important trends.  (1) The values for Main Pulses are approximately consistent with the JB value.  (2) The dispersion measures for the High-Frequency Interpulses can be larger or smaller than the JB values, and show significant scatter. The \DDM\ scatter for the High-Frequency Interpulse appears random.  We find no  correlation between \DDM\ and pulse time of arrival, either as function of rotation phase (shown in Figure \ref{fig:IP_dispersion_singles}) or clock time during the 1-hour observing run. 

A few points in Figure \ref{fig:IP_dispersion_singles} have \DDM $< 0$. We do not interpret this as inverted dispersion behavior ($ d t_{\rm p} / d \nu > 0$) in the pulsar.  Rather, it seems likely that the JB value, which is based on mean profiles, may contain a small contribution from the pulsar itself.     

\subsubsection{Systematics: all of the pulses}

\begin{figure}[htb] 
\includegraphics[width=\columnwidth]{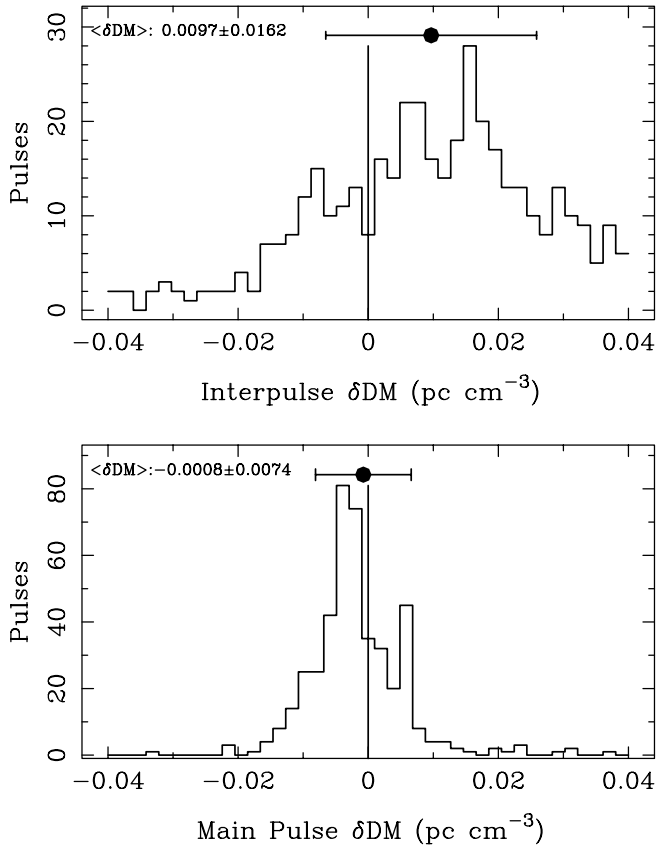}
\caption[]{Excess dispersion measures, relative to the Jodrell Bank value, for the  Interpulse (top) and the Main Pulse (bottom).   We included 356 Main Pulses and 386 Interpulses, all but 3 of which are High-Frequency Interpulses.  The pulses were captured at frequencies from 2 to 30 GHz, and are distributed over frequency as described in the caption of Figure \ref{fig:DMbyBand}. Outliers are excluded, as described in Figure \ref{fig:IP_dispersion_singles}. The spread in \DDM\ for Interpulses is clearly larger than the measurement uncertainty, and must be intrinsic to the pulsar. }
\label{fig:AveDMHist}
\end{figure}

The trends illustrated in Figure \ref{fig:IP_dispersion_singles} hold true in general, for nearly 740 pulses we captured between 2 and 30 GHz, in more than 50 observing runs spaced over 9 years.  In Figure \ref{fig:AveDMHist} we collect \DDM\ values for all of these pulses, separated by pulse type but combining all frequencies together.  This figure shows that the High-Frequency Interpulses are, on average, more dispersed than the Main Pulses and are also more dispersed than the Jodrell Bank mean value.  We find the mean \DDM\ $ \sim 0.010$ \pccm3 for the High-Frequency Interpulse, with formal standard deviation of $\sim 0.016$ \pccm3. Furthermore, there is significant scatter about this mean.  Individual pulses can have \DDM\ as large as $\sim 0.04$ \pccm3, or they can have DM less than the Jodrell Bank value (\DDM $< 0$). By contrast, there is no strong evidence for magnetospheric dispersion in the Main Pulse. The \DDM\ values for the Main Pulse are consistent with the Jodrell Bank value.

\begin{figure}[htb] 
\includegraphics[width=\columnwidth]{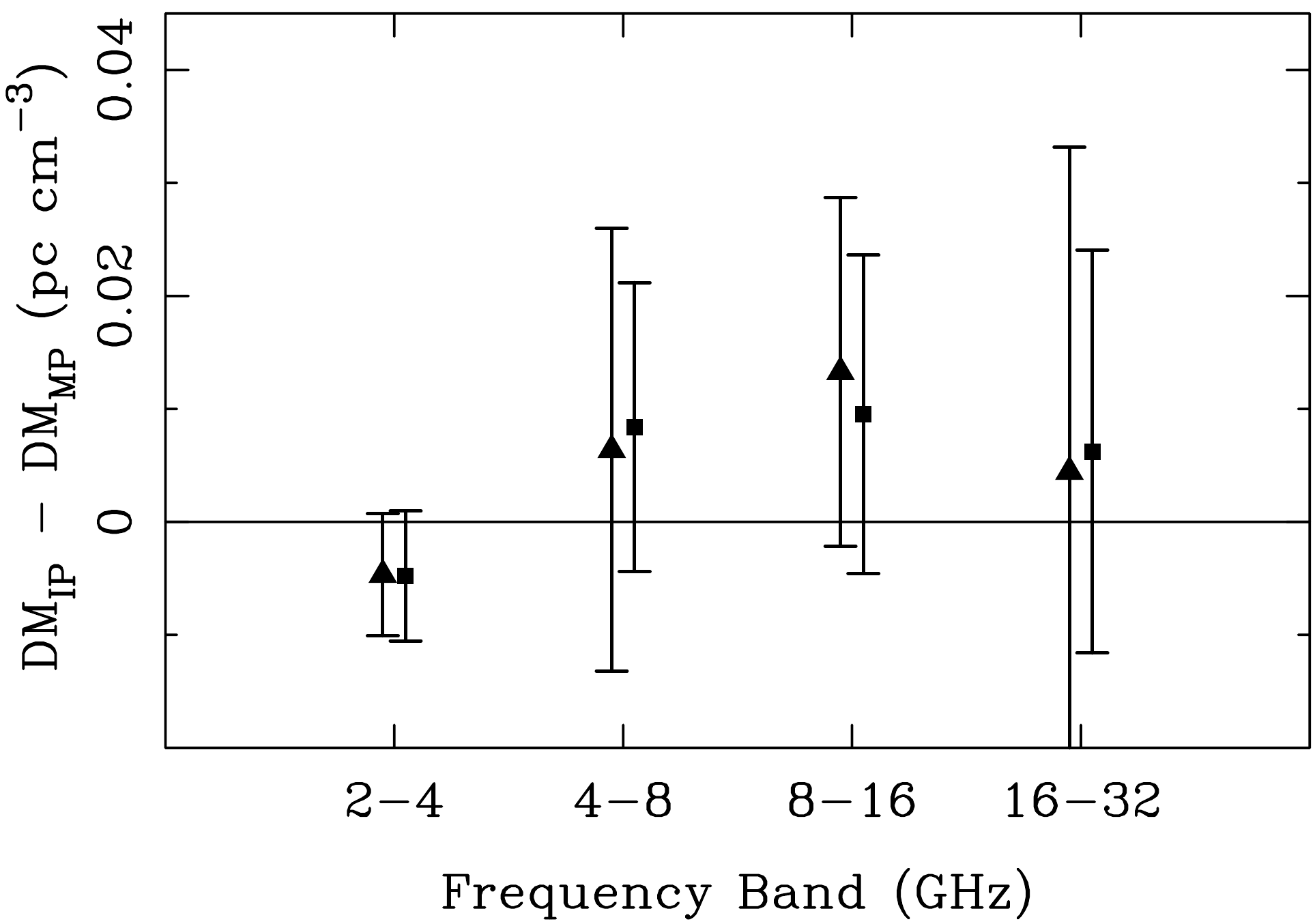} 
\caption[]{Differences in mean DM, for Main Pulses and Interpulses, binned by observing frequency and calculated by two different methods. Squares use our method 1, while triangles use method 2, as described in Section \ref{DM_analysis_details}.  Outliers have been excluded as described in Figure \ref{fig:IP_dispersion_singles}.  Bin at 2--4 GHz contains 48 MP and 3 LFIP; bin at 4--8 GHz contains 219(226) MP and 64(66) HFIP; bin at 8--16 GHz contains 80(81) MP and 217(218) HFIP; bin at 16--32 GHz contains 3(2) MP and 106(102) HFIP.  First numbers are for pulses used in method 2;  numbers in parentheses are for pulses used in method 1.  Note slight differences in outlier exclusion for the two methods. We find no strong evidence for any frequency  dependence of the excess DM for HFIPs, nor do we find any excess DM for the LFIPs.  Acronyms are given in Table \ref{table:Components}.}
\label{fig:DMbyBand}
\end{figure}

To explore the frequency dependence of the \DDM\ intrinsic to the pulsar, In Figure \ref{fig:DMbyBand} we separate the mean \DDM\ values by pulse type and by frequency.   Pulses at 2--4 GHz are Main Pulses and Low-Frequency Interpulses.  Pulses at higher frequencies are Main Pulses and High-Frequency Interpulses. This figure shows three trends. (1) There is no DM difference between Main Pulses and Low-Frequency Interpulses.   We find no evidence of intrinsic DM associated with Low-Frequency Interpulses.  (2) High-Frequency Interpulses are on average more dispersed than Main Pulses, and we find significant scatter in their excess DM.  (3) We find no evidence  of frequency dependence in the excess DM of High-Frequency Interpulses.  The excess dispersion of the High-Frequency Interpulse can plausibly be described by the $1 / \nu^2$ behavior we assumed in equation (\ref{tp_DM}). 

\subsection{Origin of the intrinsic dispersion}

High-Frequency Interpulses can be significantly more dispersed than either Main Pulses or Low-Frequency Interpulses.  We find the excess dispersion can fluctuate from zero to $\sim0.04$ \pccm3 on timescales of no more than a few minutes.  Our results constrain the nature and/or location of the emission zone for this component.

The \DDM\ scatter of the High-Frequency Interpulse cannot be caused by the Crab Nebula or the interstellar medium between us and the Crab pulsar.  In neither of these can the dispersion have changed in less than a minute (the typical time between subsequent bright pulses we captured).  Furthermore, there is no reason why dispersion from the Nebula or the interstellar medium changed in phase with the pulsarÕs rotation (which would be needed to enhance dispersion of only the High-Frequency Interpulse).  We conclude {\it the dispersion measure fluctuations are intrinsic to the magnetospheric emission region for the High-Frequency Interpulse}. 

Furthermore, the excess \DM\ cannot be from some dense region that just happens to sit above the emission region for the High-Frequency Interpulse, nor can it be due to a longer propagation path through the magnetosphere for that component,  In either case we would see large \DDM\ in every High-Frequency Interpulse -- which is not the case.  Instead, the excess \DM\ must arise locally in the emission zone, perhaps due to turbulent fluctuations in the plasma density around the emission region.  We discuss this further in \citet{EH2016} where we present one possible example of the dispersing medium.

\section{High-Frequency Components}
\label{section:HFCs}

Although High-Frequency Components (HFCs) are easy to detect in mean profiles between 3.5 and 28 GHz \citep[see Figure 1 in][]{Hankins2015}, the associated single pulses are generally fainter than Main Pulses or Interpulses, which makes them hard to observe individually.  \citet{Jessner2005} captured a few dozen at 8.35 GHz, and \citet{Mickaliger2012} reported ``a few'' HFC single pulses at 8.9 GHz, but neither paper discussed the characteristics of single High-Frequency Component pulses.  In one exceptional observing run at the Arecibo Observatory we were able to catch 30 bright pulses between 8 and 10.5 GHz.  These were evenly split between the first and second High-Frequency Components (which we refer to as HFC1 and HFC2, when needed, to shorten the notation).

\begin{figure}[htb] 
\includegraphics[width=\columnwidth]{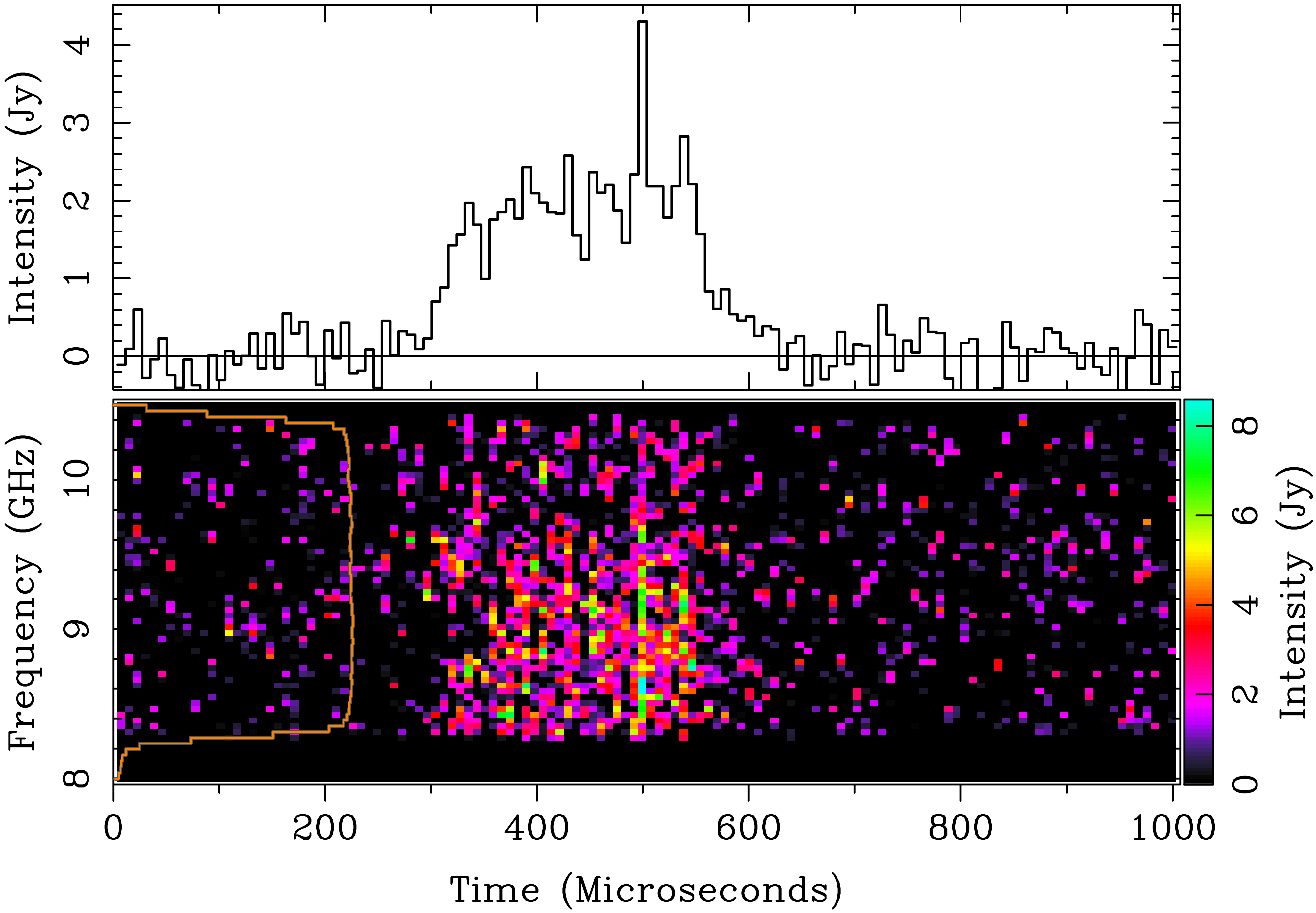}
\includegraphics[width=\columnwidth]{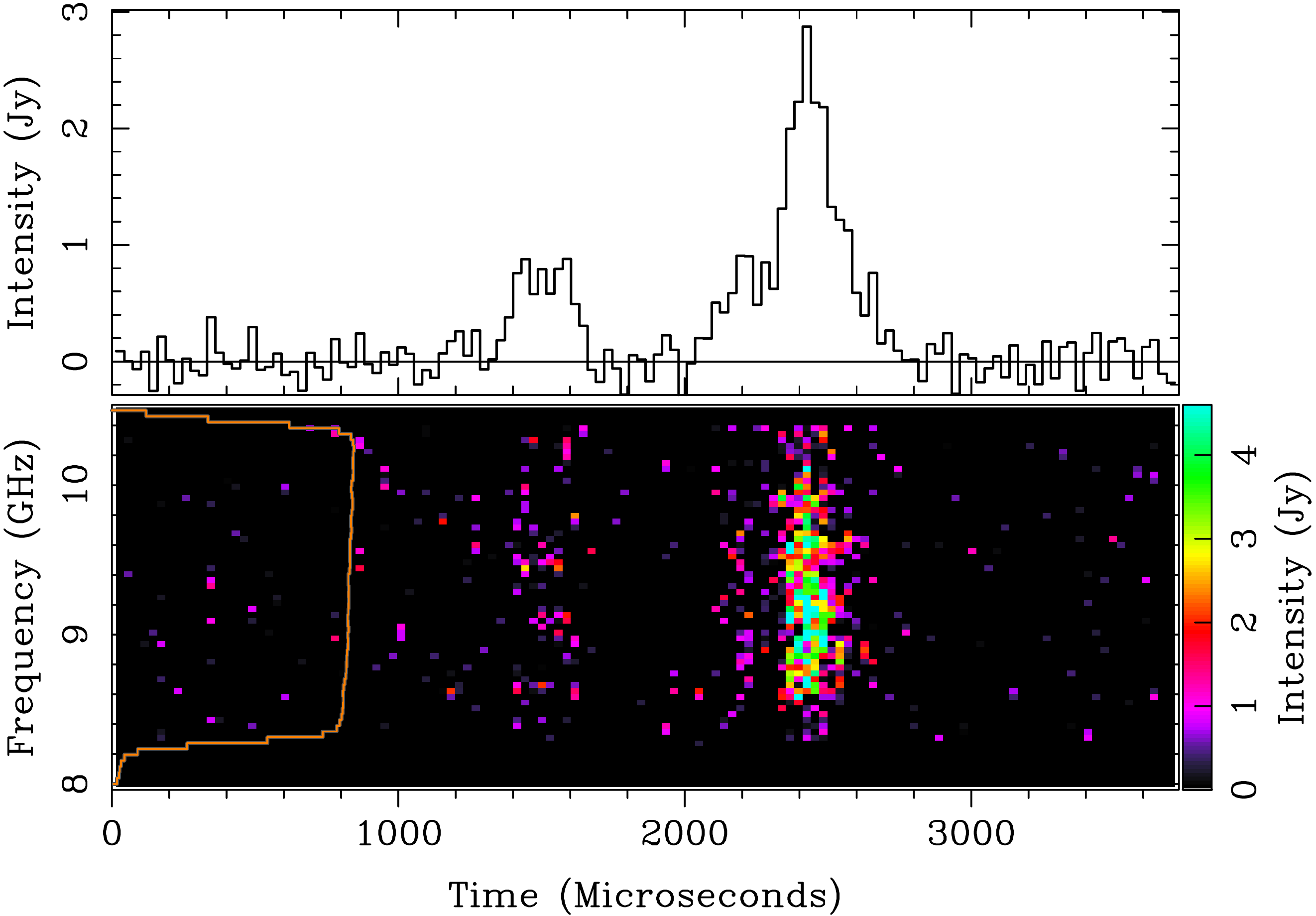}
\caption{Two examples of High-Frequency Component pulses recorded between 8 and 10.5 GHz at the Arecibo Observatory. Above, a pulse from HFC1;  below, a pulse from HFC2.  Both are plotted with a spectral resolution of 39 MHz, and de-dispersed at DM of 56.76080 \pccm3.  The time resolution of the upper pulse is 7.8 $\mu$s, and the lower pulse, 28.7 $\mu$s. These pulses last much longer than the Main Pulses shown in Figure \ref{fig:MPs_above_10GHz} and the Interpulses shown in Figure \ref{fig:IP_at_two_freqs}, but are also much weaker.
\label{fig:HFC}}
\end{figure}
  
\subsection{Single High-Frequency-Component Pulses}

The temporal and spectral characteristics of High-Frequency-Component pulses are reminiscent of Main Pulses and Low-Frequency Interpulses.  To illustrate, in Figure \ref{fig:HFC} we show the total intensity and dynamic spectrum of two example pulses.  Single High-Frequency-Component pulses -- both HFC1 and HFC2 --  can have one or a few components.  The dynamic spectrum of a High-Frequency-Component pulse is relatively broadband, extending across our 2.5 GHz observing band.   In both of these aspects  they resemble Main Pulses and Low-Frequency Interpulses.

There are differences, however. Unlike the Main Pulse and the Low-Frequency Interpulse, the High-Frequency Components show strong linear polarization.  This is evident in the pulsar's mean profile \citep{MH1999}, and is consistent with the single pulses we captured.  While many of those pulses were too weak to measure significant polarization, the stronger pulses typically showed $\sim 50\%$ linear polarization (at $3.2~ \mu$s time resolution), with approximately constant position angles.  We did not  catch any pulses with significant circular polarization or  significant position angle rotation. 

High-Frequency-Component pulses are fainter, and last longer, than Main Pulses and  Low-Frequency Interpulses.  The peak flux of the High-Frequency-Component pulses we captured was no more than a few Jy,  much fainter than the $100-1000$ Jy peak flux typical of bright Main Pulses and Interpulses at the same frequency.  This is consistent with \citet{Jessner2005}, who found the flux distribution of High-Frequency-Component pulses at 8 GHz to be much steeper than that of Interpulses.  To  measure the duration, we use the equivalent width of the intensity, as in \citet{Hankins2015}.  We find the mean equivalent width of our HFC1 pulses is $540 \pm 220~ \mu$s for HFC1 pulses and $283 \pm 102~ \mu$s for HFC2 pulses.  For comparison, the equivalent widths of single Main Pulses and High-Frequency Interpulses are typically $\sim 1-2 ~\mu$s at the same frequency \citep{Hankins2015}.

\subsection{Origin of High-Frequency Components?}

We have found little discussion in the literature on the possible origin of the High-Frequency Components.  Perhaps they come from scattering of another component, or perhaps they are separate emission phenomena. 

\subsubsection{Scattering}

It is tempting to consider the High-Frequency Components as scattered versions of a different profile component. Their broader width is reminiscent of pulse broadening in propagation through turbulent media, and their total fluence at 9 GHz is similar to that of the Main Pulses and High-Frequency Interpulse we see at the same frequency. This picture is challenged, however, by the differences between single pulses belonging to each component. Main Pulses are not strongly polarized, but High-Frequency-Component pulses (probably) show strong polarization. Main Pulses are rare above 10 GHz, where the High-Frequency Component is strong. The dynamic spectra of High-Frequency-Component pulses does not show the distinctive spectral bands characteristic  of High-Frequency Interpulses.  It is not clear that simple turbulent scattering could account for these differences. 

\citet{Pet2008,Pet2009} also proposed a scattering model, based on stimulated magnetized Compton scattering as a pulse passes through the relativistic pair plasma in the magnetosphere.  For the Crab pulsar, she argues that the Main Pulse originates at low, polar-cap altitudes, then is scattered twice, both scatterings happening at high altitude. The first scattering produces the Low-Frequency Component, which is itself then scattered to higher frequencies to produce the two High-Frequency Components.  Petrova points out that, if the details are right, the  scattered radiation can be strongly polarized and can emerge at higher frequencies than the incoming radiation. 

In addition to explaining the different characteristics of the High-Frequency Components and the pulses which are scattered to make them,  we note that any scattering model must  explain why two discrete High-Frequency Components exist -- rather than a broad range of such emission --  and why the phases of those two components change with frequency.  More work is needed here before such models can fully explain the Crab pulsar. 

\subsubsection{Separate emission phenomena}

If the High-Frequency Components are not scattered versions of some other component, they must arise from separate emission regions somewhere in the magnetosphere. Their unusual rotation phases suggest they are not low-altitude, polar cap components, but -- as with the Main Pulse and both Interpulses -- come from high magnetospheric altitudes. 

Most models of caustic emission assume the caustics connected to the two polar caps are symmetric, with uniform emission throughout the extended gap region. These models usually produce double-peaked mean profiles, but with the right choices of emissivity distribution and viewing angle they can predict more complex mean profiles \citep[e.g.,][]{BaiS20,Breed15}. It may be that the High-Frequency Components come from a ``just right'' mix of geometry and emissivity distribution within a high-altitude caustic. As a specific example,  \citet{RW10} noted that a subset of the last closed field lines can extend for a very long distance before crossing the light cylinder or the null charge surface.  They speculated that these field lines may define a disjoint, high-altitude gap surface which could be the source of the High-Frequency Components.  

If the High-Frequency Components arise from physically distinct emission regions, must a distinct emission physics also be operating in those regions? Clearly we cannot invoke the same emission mechanism as for the High-Frequency Interpulse, because we do not see spectral emission bands. Because we have not captured enough  strong pulses to comment on the existence of nanoshots, we cannot easily rule out the microscale emission mechanism that operates in the Main Pulse and the Low-Frequency Interpulse.   The longer duration of High-Frequency-Component bursts,  compared to Main Pulses and Low-Frequency Interpulses,  could be due to differences in the energy storage and/or  release mechanisms in the emission zones for each component.  The higher polarization of High-Frequency-Component pulses could, perhaps,  reflect a stronger magnetic field in their emission zones. We therefore conclude that a third type of emission mechanism may be operating in the High-Frequency-Component source regions, but one is not required.

\section{Summary and more questions}
\label{section:Last}
 In this paper we presented our new  single-pulse observations of the Crab pulsar between 10 and 43 GHz, and combined these results with our previous work to characterize each of the five emission components which dominate the pulsar's mean radio profile above 1 GHz.  Our results clarify the properties of single pulses from each of these components, but also raise new questions about radio emission sites in the Crab's magnetosphere. 

\subsection{Summary by component}

We begin by summarizing the results of our single-pulse observations of the Crab pulsar,  organized by mean-profile component. 

\subsubsection{Mean Pulse and Low-Frequency Interpulse} 

Main Pulse emission between 1 and 30 GHz comes in one or a few microbursts, each burst on the order of microseconds long. The spectrum of a microburst is relatively  broadband; it is continuous across our few-GHz observing bandwidth.  Microbursts show complex substructure; both intensity and polarization vary on timescales as short as several nanoseconds.  Very rarely the nanoshots within a Main Pulse are sparse enough to be studied individually.  The spectra of these nanosecond-long flashes of coherent radio emission is relatively narrowband, smaller than our observing bandwidth. We argue that every Main Pulse is a clump of such nanoshots, with varying central frequencies,  which are usually not resolved because they overlap in time.

Low-Frequency Interpulses show the same temporal and spectral characteristics as Main Pulses.  Even though these two components likely come from different parts of the pulsar's magnetosphere -- probably high-altitude caustics associated with the star's two magnetic poles -- we infer they involve the same emission physics.  

Both Main Pulses and Low-Frequency Interpulses become rare above $\sim 10$ GHz,  but we did capture one Main Pulse at 43 GHz.  Its spectrum and polarization are very different from the spectra of Main Pulses at lower frequencies.  A sample of one hardly justifies further discussion, but we find this one unusual pulse intriguing.

\subsubsection{ High-Frequency Interpulse}

High-Frequency Interpulses are very different from Main Pulses and Low-Frequency Interpulses.  Although they do contain microbursts,  we have never seen nanoshots in any High-Frequency Interpulse. High-Frequency Interpulses are not seen below $\sim 5$ GHz, but they continue strong and abundant to at least 30 GHz. Most importantly, the unusual emission bands in the dynamic spectrum of the High-Frequency Interpulse continue unchanged, with inter-band spacing given by  $\Delta \nu = 0.06 \nu$, over this full frequency range. Although we have not observed a single pulse over the full 25 GHz range, we hold it likely that the spectrum of a single High-Frequency Interpulse contains at least $\sim 30$ emission bands. Because these differences from the Main Pulse and Low-Frequency Interpulse are so striking, we infer that different emission physics is operating in this component.

High-Frequency Interpulses are partially dispersed within the pulsar magnetosphere. The 
intrinsic dispersion varies rapidly, changing by $\sim 0.02$ \pccm3, on timescales of a few minutes.  High-Frequency Interpulses show strong linear polarization, with a position angle that is generally constant across the pulse and independent of the rotation phase within the ``window'' defined by the component in the mean profile. These results suggest that the emission region which creates the High-Frequency Interpulse is highly dynamic, probably turbulent, with a magnetic field that has a nearly constant direction throughout the region. 

\subsubsection{High-Frequency Components}

Because High-Frequency-Component pulses are fainter than their Main Pulse and Interpulse counterparts, they are hard to observe.  We managed to catch a modest number between 8 and 10.5 GHz. These pulses are faint but long-lived, lasting several hundred microseconds. Their spectra are broadband, extending across our 2.5 GHz observing bandwidth.  They tend to be linearly polarized, with approximately constant position angles that are independent of the rotation phase of the pulse. 

If the High-Frequency Components come from separate emission regions in the magnetosphere, their longer duration and high polarization may reflect different dynamics in their natal regions.  Alternatively, it may be that they are secondary phenomena, produced by scattering or re-direction of other mean-profile components. 

\subsection{What have we learned?}

We continue by summarizing what our results reveal about radio emission physics, and radio loud regions, in the magnetosphere of the Crab pulsar.  

\vspace{2ex}
$\bullet$ {\em At least two types of radio emission physics operate in the magnetosphere of the Crab pulsar.} 

The Main Pulse and the Low-Frequency Interpulse are characterized by {\em nanoshot emission}: narrow-band, nanosecond-long flashes of emission, usually clumped together in microbursts. This emission is strong at low radio frequencies, but dies out above several GHz. Because each nanoshot can have its own polarization signature, this emission type can be depolarized in mean profiles.

The High-Frequency Interpulse is characterized by {\em spectral band emission:} longer bursts of strongly polarized emission containing of distinctive spectral emission bands.  This emission is strong at high radio frequencies but dies out below 4 GHz. The steady  position angle of the linear polarization means this emission type is strongly polarized in mean profiles.

We suspect the High-Frequency Components may be  a third type of radio emission, but we do not have enough data to characterize them well.  We have no high-time-resolution data on individual Precursor or Low-Frequency Component pulses, so cannot comment on their radio emission physics. 

\vspace{2ex}
$\bullet$ {\em There are several radio emission sites in the magnetosphere of the Crab pulsar.}

The similar radio characteristics of the Main Pulse and the Low-Frequency Interpulse, and their phase coincidence with high-energy pulses, suggests the emission regions for these two components lie somewhere along the high-altitude caustics above each of the star's two magnetic poles.  

The rotation phases of the Precursor and the Low-Frequency Component suggest those components come from low altitudes.  Perhaps these two components arise close to one of the pulsar's polar caps, in accordance with standard models of rotation-powered pulsars.

The different radio characteristics of the High-Frequency Interpulse suggest its emission region is physically separate from that for the Low-Frequency Interpulse, despite the similar rotation phases of those components.  The strong polarization and fluctuating dispersion of the High-Frequency Interpulse suggest its emission zone is a spatially localized, turbulent region. 

The unusual phases of the two High-Frequency Components suggest yet more radio emission zones -- or localized scattering sites -- exist somewhere in the magnetosphere.  The variable phases of these  components suggests that the location of these emission or scattering sites changes with observed radio frequency.

Because the structure of the extended caustics is still poorly understood, we are not able to constrain where in the magnetosphere the emission zones for the various high-altitude radio components might be.  Perhaps radio data such as ours can guide future modeling of the density and magnetic field structures within the caustics.  

\subsection{What mysteries remain?}

To conclude, we note that our observations have raised as many questions as they have answered.  We suggest a few such questions here;  no doubt the reader can think of others.

\begin{itemize} 
\item Is there an unseen, High-Frequency Main Pulse in the Crab pulsar?  If the High-Frequency Interpulse comes from a separate part of a caustic that gives rise to the Low-Frequency Interpulse, perhaps the caustic that gives rise to the Main Pulse also has such a region, which we have not yet detected.

\item Do the main radio emission mechanisms in the Crab pulsar -- nanoshot emission and spectral-band emission -- occur in other pulsars?  How does low-altitude, polar cap emission compare to high-altitude emission in the Crab pulsar? 

\item What physical conditions trigger a particular type of radio emission?  For instance, is the trigger related to density, magnetic field, rotation rate, and/or  particle acceleration in gap regions? Or is the trigger due to something totally different?

\item Why do only a few pulsars show complex, high-altitude radio emission that is phase-aligned with high-energy pulsed emission?  Is there a fundamental difference in the high-altitude magnetospheres of these objects, or do many pulsars have high-altitude radio emission -- perhaps at higher radio frequencies -- which has not yet been found?
\end{itemize}

We hope that our detailed study of the Crab pulsar can inspire similar observations of other pulsars, and future modeling of likely magnetospheric emission zones, that may answer some of these questions. 

\begin{acknowledgements}
We thank the technical, operations, and computer staffs at the Green Bank Telescope, the Very Large Array and the Arecibo Observatory for their help with the data acquisition equipment, observing, and for providing some of the computing environment we used for the observations.  We thank Jared Crossley, Jeff Kern, David Moffett, and James Sheckard for help with the Arecibo observations.  We thank an anonymous referee for helpful comments which have strengthened the paper. This work was partially supported by NSF grant AST-0607492.
\end{acknowledgements}

\appendix
\section{Noise and Trigger Levels}

We summarize our observing parameters in Table \ref{table:Noise_Levels}. Intensity calibrations were performed using standard calibration sources.  Differences in noise levels depend upon total receiver bandwidths, spectral resolution, detector smoothing time constant, telescopes, and at the higher frequencies, weather. The combination of the Crab Nebula spectral index and frequency dependent antenna beam widths causes the contribution from the Crab Nebula to vary greatly over the frequency range we observed. The narrow synthesized beam of the VLA resolves most of the Nebula and hence has far lower on-source system temperatures at low frequencies than the Arecibo telescope and the Green Bank Telescope where the system temperatures are completely dominated by the Nebular contribution, even at the highest frequencies we used.
 
 The integration time constant for the square-law detected trigger channel was chosen to match the dispersion sweep time across the trigger channel bandwidth.  In most cases the Trigger Level (column 10) is lower than the Intensity Noise Level (column 11) because the Trigger Detector Time Constant (column 9) is much longer than the time resolution of the dedispersed intensity shown in the figures. For figures in which we show both the total intensity and the dynamic spectrum the noise levels in columns 11 and 12 of Table \ref{table:Noise_Levels} reflect the full-band noise level and the noise level for an individual spectral channel. The noise levels and sensitivity of the GAVRT telescope over the 2.5 to 10.5 GHz frequency range (Figure \ref{fig:GAVRT_pulse}) are more complex.  The on-source $T_{\rm sys}$ is $\sim140$ K at 2 GHz, 60 K at 7 GHz, and 45 K at 10 GHz \citep[see Figure 4.11,][]{Jones2009}. The GAVRT telescope sensitivity ranges between 0.65 and 0.83 K/Jy \citep{Jones2010}.
\onecolumngrid

\begin{deluxetable}{cccccccccccc}
\tablewidth{\textwidth}
\tablecaption{Off-Pulse Noise and Trigger Levels \label{table:Noise_Levels}}
\tablehead{
\colhead{Figure}    &\colhead{Freq.}&\colhead{Observation}&\colhead{Telescope} &\colhead{On}           &\colhead{Sensi-}          &\colhead{Data}     &\colhead{Trigger}  &\colhead{Trigger} &\colhead{Trigger}&\colhead{Intensity}&\colhead{Spectrum} \\
\colhead{in}        &\colhead{(GHz)}&\colhead{Date}       &\colhead{ }         &\colhead{Source}       &\colhead{tivity}          &\colhead{Channel}  &\colhead{Channel}  &\colhead{Detector}&\colhead{Level}  &\colhead{Noise}    &\colhead{Noise}    \\
\colhead{text}      &               &\colhead{ }          &\colhead{ }         &\colhead{$T_{\rm sys}$}&\colhead{(K/Jy)}          &\colhead{Band-}    &\colhead{Band-}    &\colhead{Time-}   &\colhead{(Jy)}   &\colhead{Level}    &\colhead{Level}    \\
\colhead{ }         &\colhead{ }    &\colhead{ }          &\colhead{ }         &\colhead{(K)}          &\colhead{ }               &\colhead{width}    &\colhead{width}    &\colhead{constant}&\colhead{ }      &\colhead{(Jy)}     &\colhead{(Jy)}     \\
\colhead{ }         &\colhead{ }    &\colhead{ }          &\colhead{ }         &\colhead{ }            &\colhead{ }               &\colhead{(MHz)}    &\colhead{(MHz)}    &\colhead{($\mu$s)}&\colhead{ }      &\colhead{ }        &\colhead{ }               
} 
\tablecolumns{11}
\startdata
\phn1a              & \phn1.38      &1998/12/27           &VLA\tablenotemark{a}&140                    &\,1.4\tablenotemark{d}\phn&\phn\phn50         &\phn\phn\phn0.25    &\phn40           &270\phd\phn      &38\phd\phn\phn    & 130                \\
\phn1b              & \phn1.42      &1994/02/23           &VLA\tablenotemark{ }&140                    &2.5\tablenotemark{ }\phn  &\phn\phn50         &\phn\phn\phn0.25    &\phn40           &150\phd\phn      &20\phd\phn\phn    & \phn66             \\
\phn3a              &    14.00      &2010/12/02           &GBT\tablenotemark{b}&100                    &1.5\tablenotemark{ }\phn  &4000               &\phn450\phd\phn\phn &\phn78           &\phn\phn3.0      &14\phd\phn\phn    & \phn73             \\ 
\phn3b              &    20.25      &2009/04/13           &GBT\tablenotemark{ }&\phn56                 &1.0\tablenotemark{ }\phn  &4260               &\phn450\phd\phn\phn &\phn25           &\phn\phn4.5      &\phn7.2\phn       & \phn37             \\ 
\phn4\phn           & \phn9.25      &2005/01/05           &AO\tablenotemark{c} &102                    &4.0\tablenotemark{ }\phn  &2220               &\phn140\phd\phn\phn &\phn83           &\phn\phn2.0      &\phn3.4\phn       &                    \\  
\phn\phn5, 6\phn\phn& \phn9.25      &2007/01/01           &AO\tablenotemark{ } &102                    &4.0\tablenotemark{ }\phn  &2200               &\phn140\phd\phn\phn &\phn83           &\phn\phn2.1      &13\phd\phn\phn    &                    \\ 
\phn7a              & \phn9.25      &2005/08/08           &AO\tablenotemark{ } &\phn65                 &4.0\tablenotemark{ }\phn  &2200               &\phn140\phd\phn\phn &\phn83           &\phn\phn1.3      &\phn2.1\phn       & \phn15             \\ 
\phn7b              & \phn7.00      &2006/03/11           &AO\tablenotemark{ } &168                    &5.0\tablenotemark{ }\phn  &1070               &\phn\phn50\phd\phn\phn &\phn70        &\phn\phn4.8      &\phn5.9\phn       & \phn22             \\ 
\phn8\phn           & \phn9.25      &2005/08/08           &AO\tablenotemark{ } &\phn65                 &4.0\tablenotemark{ }\phn  &2200               &\phn140\phd\phn\phn &\phn83           &\phn\phn1.3      &\phn7.3\phn       &                    \\ 
\phn9\phn           &    43.25      &2009/04/18           &GBT\tablenotemark{ }&\phn70                 &0.46\tablenotemark{ }     &4060               &1000\phd\phn\phn    &\phn20           &\phn\phn7.9      &\phn9.6\phn       & \phn92             \\
10a                 &    14.00      &2010/12/02           &GBT\tablenotemark{ }&100                    &1.5\tablenotemark{ }\phn  &4000               &\phn450\phd\phn\phn &\phn77           &\phn\phn3.0      &\phn9.9\phn       & \phn75             \\ 
10b                 &    28.00      &2010/12/05           &GBT\tablenotemark{ }&\phn82                 &1.0\tablenotemark{ }\phn  &4450               &1000\phd\phn\phn    &\phn22           &\phn\phn7.0      &15\phd\phn\phn    & 110                \\ 
12a                 &    20.25      &2009/04/13           &GBT\tablenotemark{ }&\phn56                 &1.0\tablenotemark{ }\phn  &4250               &\phn450\phd\phn\phn &\phn25           &\phn\phn4.5      &                  &                    \\ 
12b                 &    14.00      &2010/12/02           &GBT\tablenotemark{ }&100                    &1.5\tablenotemark{ }\phn  &4000               &\phn450\phd\phn\phn &\phn77           &\phn\phn3.0      &                  &                    \\ 
13, 14              &    20.25      &2009/04/12           &GBT\tablenotemark{ }&\phn56                 &1.0\tablenotemark{ }\phn  &4250               &\phn450\phd\phn\phn &\phn25           &\phn\phn4.5      &3.6\phn           &                    \\ 
15\phn              &    20.00      &2010/12/02           &GBT\tablenotemark{ }&101                    &1.0\tablenotemark{ }\phn  &4650               &\phn450\phd\phn\phn &\phn26           &\phn\phn7.9      &9.6\phn           &                    \\ 
19a                 & \phn9.25      &2008/04/13           &AO\tablenotemark{ } &101                    &4.0\tablenotemark{ }\phn  &2070               &\phn450\phd\phn\phn &270              &\phn\phn0.6      &0.3\phn           & \phn\phn1          \\
19b                 & \phn9.25      &2008/04/13           &AO\tablenotemark{ } &101                    &4.0\tablenotemark{ }\phn  &2070               &\phn450\phd\phn\phn &270              &\phn\phn0.6      &0.14              & \phn\phn\phn1   
\enddata
 \tablecomments{a. Very Large Array. b. Green Bank Telescope. c. Arecibo Observatory. d. Only half of the VLA antennas were used for this observation.\\
}
 \end{deluxetable}

\end{document}